\documentclass[]{aa}

\usepackage{physics}
\usepackage[]{graphicx}
\usepackage{subcaption}
\usepackage{tabularx}
\usepackage{ gensymb }
\usepackage{ textcomp }
\usepackage{txfonts}
\newcommand{\Msol}{M$_\odot$}
\newcommand{\citeg}[1]{\citep[e.g.,][]{#1}}

\begin{document}

   \title {Dependence of the dynamical properties of light-cone simulation dark matter halos on their environment}
   \titlerunning{Dependence of dark matter halo properties on the environment}

   \author{Maria Chira
          \inst{1,2}\thanks{mchira@physics.auth.gr}
          \and
          Manolis Plionis\inst{2,1}
          \and
          Shankar Agarwal\inst{3}
          }

   \authorrunning{Chira
          \and
          Plionis
          \and
          Agarwal
          }

   \institute{Physics Dept., Aristotle Univ. of Thessaloniki, Thessaloniki
  54124, Greece
         \and National Observatory of Athens, P.Pendeli, Athens, Greece
         \and African Institute for Mathematical Sciences, 6 Melrose Road, Muizenberg 7945, Cape Town, South Africa
             }

   \date{Received ; accepted }


  \abstract
   {}
   {We study the dependence of the dynamical properties of massive $(M \geq 1.5\times10^{13} M_{\odot}/h)$ dark matter (DM) halos on their environment in a whole-sky $\Lambda$CDM light-cone simulation extending to $z\sim 0.65$. The properties of interest for this study are the halo shape (parametrized via its principal axes), spin and virialization status, the alignment of halo spin and shape, as well as the shape-shape and spin-spin alignments among halo neighbors. }
   {We define the halo environment using the notion of halo isolation status determined by the distance to its nearest neighbor. This defines a maximum spherical region around each halo devoid of other halos, above the catalog threshold mass. We consider "close halo pairs" to be pairs  separated by a distance that is lower than a specific threshold. In order to decontaminate our results from the known dependence of halo dynamical properties on mass, we used a random sampling procedure to compare properties of similar halo abundance distributions.}
   {We find that: (a) there is a strong dependence on the part of the halo properties with regard to their environment, confirming that isolated halos are more aspherical and more prolate with lower spin values; (b) correlations between halo properties exist and are mostly independent of the halo environment; (c) halo spins are aligned with the minor axis, regardless of halo shape; and (d) close halo neighbors have their major axes statistically aligned, while they show a slight but statistically significant preference for anti-parallel spin directions. The latter result is enhanced for the case of close halo pairs in low-density environments. Furthermore, we find a tendency for the spin vectors to be oriented perpendicular to the line that connects such close halo pairs.}
   {}


   \maketitle
%

\section{Introduction}

According to the cold dark matter (CDM) cosmological model, structure in the universe forms hierarchically. Dark matter (DM) perturbations of different sizes, following the CDM power spectrum, grow to eventually form DM halos, inside of which baryonic matter collapses to form the visible structure made up of galaxies, stars, etc \citep{White1978,White1991}. The result of such a structure formation process is what is known as the "Cosmic Web" \citep{Bond1996,RVdW16}, consisting of four main structure elements: knots, filaments, sheets, and voids.

Since galaxy formation occurs in the inner regions of DM halos, galactic properties are expected to be related to those of their host DM halo. A number of authors have investigated galactic and halo properties using observational surveys and N-body simulations to understand the physical processes that lead to the formation of cosmic structure on a variety of scales.

 An important question is focused on how halos acquire their shapes and angular momenta. According to the tidal alignment theory, the shape of DM halos is mainly the result of the anisotropic collapse of matter in directions determined by the gravitational tidal field \citep{West1995, Catelan2001}. Moreover, according to the tidal torque theory \citep[TTT,][]{Hoyle1951,Peebles1969,Doroshkevich1970,White1984}, it is the misalignment between the shape tensor of the halo and the shear field tensor that leads to the acquisition of angular momentum up to the turnaround point, when collapse becomes non-linear and the effect of the tidal field is gradually eliminated. From this point on, the angular momenta of halos and galaxies are affected by non-linear effects and baryonic physics, which collectively lead to a mean misalignment of~50\textdegree of the angular momenta axes predicted by the TTT and those found in N-body simulations for the present epoch \citep{Porciani2002a}. The latter serves as the most appropriate tool for following structure formation during the non-linear collapse stage \citep{Barnes1987, Porciani2002b, Sugerman2000,2000Colberg, Lopez2019}.

It has been shown that the shape of non-linear cosmic structure is an important ingredient for understanding structure formation processes and a substantial effort has been devoted to quantifying observationally the shape of non-linear structures. Observations centered on the kinematics of elliptical galaxies and analyses of early N-body simulations \citeg{Hohl1971} have led to a revisiting of notions of the shape of galaxies and halos \citeg{Contopoulos1956,Stark1977,Binney1976}, which had previously been considered to be axially symmetric and particularly oblate.  Ever since the first relatively high-resolution simulations became available, the shape distributions of DM halos began to be rigorously studied and it is a common prediction that halos are triaxial, with the majority of authors arguing for prolate halo shapes \citeg{Frenk1988,Warren1992,Faltenbacher2002,Shaw2006,Allgood2006,Plionis2006,Hahn2007a,Bett2007,VegaFerrero2017,GaneshaiahVeena2018}. This preference for prolateness has been found to be more prominent for massive halos \citep{Shaw2006,Allgood2006,AvilaReese2005,Gottlober2006,Bett2007,VegaFerrero2017,Bailin2005}, which seems to be a consequence of the hierarchical formation process: massive halos are known to have formed more recently and their shape is more affected by their last major merging events within the supercluster network than that of low-mass halos at the present epoch \citep{vanHaarlem1997,Vitvitska2002,2000Colberg,Porciani2002b,VegaFerrero2017}.

Concerning the acquisition of angular momentum and its evolution, predictions of the first triaxial galaxy models \citeg{Schwarzschild1979} already indicated that the streaming motions of stars are allowed around the galactic short and long axis and that streaming around the intermediate axis corresponds to an unstable configuration \citep{Heiligman1979}. \citet{Shaw2006} stated that rotation around the minor axis is the most stable configuration and, thus, this is what would be expected to be observed or found in cosmological N-body simulations. Indeed, \citet{Franx1991} find that most of the 38 elliptical galaxies in their sample have their rotation axes well-aligned with their short axes, although the range of misalignment angles are between 0 and 90 \textdegree. Using cosmological N-body simulations, a plethora of authors have confirmed a tendency of the angular momentum of halos to align with the minor axis \citep{Barnes1987,Dubinski1992,Warren1992,Bailin2005,Allgood2006,AvilaReese2005,Shaw2006,Bett2007,GaneshaiahVeena2018}. However, the width of distribution of misalignment angles is large, while the corresponding distribution for the intermediate axis is found to be close to uniform, in accordance with the theoretical models. \citet{Shaw2006} separated their halo sample into subsamples of spin parameter bins to find that halos with high spin values tend to align better their spin with their minor axis, while the scatter of the alignment distribution decreases with increasing spin magnitude. They also state that the deformation-spin relation of \citet{Peebles1969} does not hold for the halos in their simulations and spin cannot be the cause of halo shape. However, the relative alignment of angular momentum with the halo shape indicates, at least partially, a common origin for the halo shape and spin \citep{Allgood2006}.

The environmental dependence of the dynamical properties of non-linear cosmic structures ought to be better understood since it is the basis of structure formation theories, especially of the TTT \citep[for a review]{Shafer2009} that says that such properties (eg., shape and spin) are correlated with their environment, implying that halos formed under the effects of the same tidal field should preserve some  properties in common, reflecting their large-scale environment. A pioneer in the field, \citet{Binggeli1982} revealed the tendency of 44 Abell galaxy clusters to "point to each other,"\ that is,  alignment of their major axes up to separations of $~ 30$ Mpc. Following works failed to reveal any cluster alignment signal \citeg{Struble1985, Rhee1987, Ulmer1989}. However, later on, various authors have confirmed alignments of neighboring clusters using data from survey catalogs \citeg{ Lambas1990, Plionis1994, West1995, Chambers2002} and N-body simulations \citeg{Faltenbacher2002, Kasun2005}.

Several authors have approached some of these issues from a different point of view: instead of focusing on the correlation between the properties of neighboring halos, they investigate how certain properties correlate to their environment. Using a wide range of different methods for the classification of the halo environment and different data sets, they attempt to extract dependencies of halo properties with the larger or the local-scale environment. \citet{Lemson1999} conclude that no halo property, including spin and shape, except for the halo mass distribution, depends on the environment defined by local overdensity. Regarding halo spin (or halo angular momentum) magnitude, \citet{AvilaReese2005} found that halos residing in high-density regions, defined by the authors as {\em CLUSTER} halos, have lower spin values compared to halos in low-density regions, which they define as {\em VOID} halos, and galaxy-size parent halos, which they define as {\em FIELD} halos. \citet{Maccio2007} find no correlation with the environment; \citet{Hahn2007a} find that void-halos have lower spin parameter; while \citet{Bett2007, Wang2011, Lee2017} find that spin drops with local density. Regarding the halo shape, \citet{Hahn2007a} find that massive halos tend to be more prolate in clusters; whereas \citet{Lee2017} find that the elongation of halos increases in low-density regions.

As for the spin vector orientation relatively to the large-scale structure, many authors find that the spin of low-mass halos is aligned with their environment, that is, along the filament ridge or on the sheet-plane, while high-mass halos tend to have spin-vectors perpendicular to their environment. The apparent orientation "transition" happens at the "spin-flip" mass, which is found to be of the order of $10^{11}-10^{12}$ h$^{-1}$ \Msol \citeg{ Navarro2004, Bailin2005, AragonCalvo2007, Hahn2007a, GaneshaiahVeena2018,Lee2020}. Moreover, \citet{Brunino2007} and \citet{Cuesta2008} find that spin vectors tend to lie on the surface of voids.

The different spin orientations with respect to the large-scale filaments, between the low- and high-mass halos, has been established by a plethora of relative studies \citeg{Veena2020,AragonCalvo2014, Trowland2013,Codis2012,Hahn2007b,Bailin2005}, yet the explanation is unclear, as it has been vividly demonstrated in the relevant discussion of \citet{GaneshaiahVeena2018}. One interpretation that has been suggested is that mass flows towards the filamentary intersection of sheets perpendicularly to the filament ridge and, thus, low-mass halos align their spins with the filaments that they reside in, whereas massive halos are built up from mergers along the filaments and as a result, they align their angular momentum perpendicularly to the filament. However, \citet{Trowland2013} find that at high redshifts, halos -- regardless of their mass -- tend to have their spins oriented perpendicular to filaments due to initial tidal torquing; at lower redshifts, low-mass halos flip their spins to become parallel to the filament direction and there is no clear explanation for this. Such a contradictory formation history of the low-$z$ correlation between spin-orientation and halo mass as well as the lack of clarity in the results of previous works indicate that the detailed picture of structure formation and angular momentum acquisition is rather incomplete -- or that at least it requires an improved understanding.

In this work, we revisit the dynamical properties of halos using a light-cone simulation of a $\Lambda CDM$ cosmology, seeking to reveal their dependence on the environment and possible interrelations among them using a simple and clear-cut environmental criterion, defined originally in \citet{Chira2018}. \citet{Libeskind2018} has shown that an important methodological issue in studies of cosmic structure properties with respect to their environment is the quantification of the environment itself, which can have an important impact on the results. Here, our environmental criterion is based on the distance of halos and halo pairs to their nearest neighbor. Using this simple and straightforward definition, we study the dependence of halo shape, spin and virialization status, as well as the alignment of shape and spin vectors between halo neighbors and close halo pairs found in different environments. Both the selection of light-cone data and our definition of the environment render our results relatively easily comparable to observational data from current and future cluster missions such as {\rm eRosita} \citeg{Hofmann2017}, LSST \citeg{Marhall2017} and Euclid \citeg{Sartoris2016}.

\section{Simulation data}

For our study we use the light-cone halo catalogs from a subset of N-body simulations from the "Dark
Energy Universe Simulation" (DEUS) project \citep{Alimi2010, Rasera2010, Courtin2011} and publicly available through
the DEUS database\footnote{www.deus-consortium.org/deus-data/}. The
N-body runs have been performed using the adaptive mesh refinement
code RAMSES based on a multigrid Poisson solver \citep{Teyssier2002, Guillet2011} for Gaussian initial conditions
generated using the Zel’dovich approximation with the MPGRAFIC code
\citep{Prunet2008} and input linear power spectrum from CAMB \citep{Lewis2000}. The light-cone data used here are from
simulations of 2592 Mpc/h boxlength with $2048^3$ particles for a standard
$\Lambda$CDM model
with parameters calibrated against Supernova Type Ia
from the UNION dataset \citep{Kowalski2008} and measurements of the
Cosmic Microwave Background anisotropies from the Wilkinson Microwave
Anisotropy Probe (WMAP) five-year data \citep{Komatsu2009}, i.e.,
$\Omega_m=0.267$, $H_0=100 h$ km s$^{-1}$ Mpc$^{-1}$, $h=0.72$. The light-cone is built on-the-fly by storing N-body particles at each coarse time step, the particles being sampled at the appropriate comoving distance from the observer.

\begin{figure}
\centering
\includegraphics[width=0.55\textwidth]{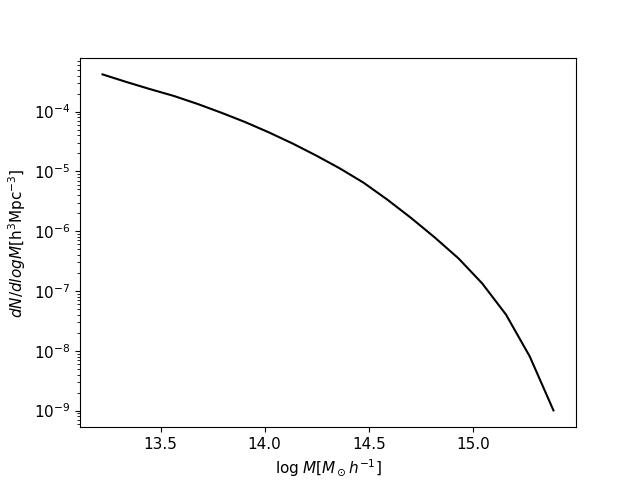}
\caption{ Mass function of the light-cone halo catalog.}
\label{MF}
\end{figure}

The light-cone halo catalogs cover the full sky across the redshift
interval, $0<z<0.65,$ with halos containing more than 100 particles and
a particle mass resolution of $m_p=1.5\times
10^{11}\,M_{\odot}/h$. Halos were detected in the light-cone using a friends-of-friends(FOF) halo finder \citep[pFOF,][]{Roy2014}. The total number of halos in our catalog is $\sim
3.15\times 10^6$ in the mass range $1.5\times10^{13}\frac{M_{\odot}}{h}\leq M \leq 3.\times10^{15}\frac{M_{\odot}}{h}$, where M is the FOF mass for linkage length $b=0.2$. The mass function of the halo-catalog is presented in Figure \ref{MF}.

\section{Method}
\subsection{Definition of local environment}\label{environment}
We use a rather simple approach to define the local
environment of dark matter halos, especially tailored to reveal the
inter-halo dynamics based on a criterion
centred on each individual halo.
We identify the distance to the nearest neighbor of
each halo, which we tag as the "isolation"
radius, $R_{\rm isol}$. Our criterion resembles that of \citet{Haas2012} although it has a major difference in that we
define the distance to any halo with mass $M\ge 1.5 \times 10^{13} M_{\odot}/h$.
We also note that although our halo catalog contains halos with $M\ge
1.5 \times 10^{13} M_{\odot}/h$, we set a lower-mass limit of $M\ge 2.5 \times 10^{13} M_{\odot}/h$ for the selection of  central halos (CHs), that is, the halos of which the nearest neighbor (of any halo mass) is used to define their local environment. This is required in order to be able to define the "isolation" of CHs towards both lower- and higher- mass halos and, thus, ensure that our results are not biased by the lower mass limit of the FOF halos.
The values of $R_{\rm isol}$ span from $\sim$0.85 to 37 $h^{-1}$ Mpc, indicating the existence of extremely isolated halos residing in huge underdense regions.
On the other extreme, the majority of halos with small $R_{\rm isol}$ should be considered as residing in high density regions. It should be noted that throughout our analysis, in order to take into account the halo size in order to categorize its local environment, we use as a criterion the isolation
radius in units of the halo virial radius, $r_{\rm vir}$, that is,
$R_{\rm isol}/r_{\rm vir}$.
Finally, as close pairs, we consider those halos that are separated by a distance, $R_{\rm isol}/r_{\rm vir}<4$. We found that close pairs of halos are encountered not just in dense environments: their isolation, quantified via the normalized distance of the CH to its second nearest neighbor, that is, $R_{\rm isol_{2}}/r_{\rm vir}$, spans a wide range extending up to $\gtrsim 30$ h$^{-1}$ Mpc \citep[see][for a detailed discussion on our definition of the environment]{Chira2018}.

We note that our use of a light-cone (which resembles the observable universe, i.e., observations at larger distances correspond to higher redshifts) instead of a fixed-redshift simulation box, as well as our definition of the halo environment are both motivated by an attempt for a more realistic setup. This is an important first step not only for a comparison of our results with the results of similar analyses on real cluster data in the future, but also for a statistical estimate of the implications of the effects studied in the current work (e.g., shape and spin alignments) on the results of weak lensing-based techniques.

\subsection{Dynamical DM halo properties}\label{method-properties}

Our main goal in this paper is to investigate if and how the dynamical state of DM halos depends on halo isolation and, when possible, to compare our results with other works, some of which use more elaborate and multiparametric characterizations of the environment.

The halo properties that we examine as a function of isolation are the following:

\vspace{0.2cm}
\noindent
{\it Halo shape} quantified by using the symmetric mass-distribution tensor \citep[also described in][]{Agarwal2015} as:
\begin{equation}
 M_{\alpha\beta}= \frac{1}{N_{h}}\sum_{i=1}^{N_{h}}(r_{\alpha i}-r_{\alpha c})(r_{\beta i}-r_{\beta c}) ,
\end{equation}
where $N_{h}$ is the number of halo particles, $\alpha,\beta=1,2,3$ correspond to the three position components. The eigenvalues, $a^{2},b^{2},c{^2}$ of the tensor $M_{\alpha\beta}$ define the fitted triaxial ellipsoid with prime axes of length $a\geq b\geq c$, while its eigenvectors denote the axes orientations. Using $a,b,c$ we investigate halo shapes in terms of flatness, which is the ratio $c/a$ of the minor over the major axis of the fitted ellipsoid (as a first parameterization of halo shape, in terms of deviation from sphericity), and triaxiality defined as:
    \begin{equation}
    T = \frac{a^2 - b^2}{a^2 - c^2}\;,
    \end{equation}
    as a more sophisticated parameterization of shape, in terms of deviation from pure spheroidal shape (oblate or prolate). The meaning of the values of T-parameter is provided in \citet{Franx1991}, shown here in Table~\ref{table:franx}. We note that in addition to flatness and triaxiality, we analyzed our data in terms of halo prolateness
    \begin{equation}
    p = \frac{a-2b+c}{2(a+b+c)}\;,
    \end{equation}
    \citep[see][]{Agarwal2015}. We find that our results are qualitatively the same, regardless of the shape parameterization being $T$ or $p$. As such, in this work, we show results only in terms of flatness and trixiality.

\vspace{0.15cm}
\noindent
{\it Spin of the prime axis} expressed by the dimensionless spin parameter,
\begin{equation}
    \lambda=\frac{J}{\sqrt{2} M_{\rm vir} R_{\rm vir} V_{\rm vir}}\;,
\end{equation}
 where $J$ is the magnitude of the angular momentum and $M_{\rm vir}, R_{\rm vir}, V_{\rm vir}$ are the virial mass, radius and velocity  of the halo respectively \citep{Bullock2001}.

\vspace{0.15cm}
\noindent
{\it Virialization status:} roughly estimated by comparing the ratio $\abs{U/K}$ of the Potential over the Kinetic energy terms with the theoretical virial value, predicted by the {\it virial theorem}, being $\abs{U/K}= 2$. We approach this parameterization with caution since the way of calculating the potential energy of DM halos in simulations leads to an underestimation of its real value. Specifically, the potential energy of a halo is calculated as the sum of the potential energy of all the halo particles. However, the potential energy of each particle is calculated by taking into account only the gravitational interaction with the halo particles and not with the non-halo particles located in the local environment of the halo. Thus, the calculated value of $U$ is only a lower limit of its actual value and, subsequently, $\abs{U/K}$ is systematically underestimated. We note that this underestimation is minimal for low-density regions where there is a smaller number of particles that are not taken into account.

\begin{table}
\caption{Triaxiality parameter for four representative ellipsoid shapes \citep[as in][]{Franx1991}}             
\label{table:franx}      
\centering                          
\begin{tabular}{c c c c}        
\hline\hline                 
$e_1 = 1 - c/a$ & $e_2 = 1 - b/a$ & $T = \frac{a^2 - b^2}{a^2-c^2}$ & Shape \\    
\hline                        
    0.5 & 0 & 0     & Oblate \\
   0.5 & 0.1 & 0.25 & Oblate-triaxial \\      
   0.5 & 0.34 & 0.75   & Prolate-triaxial \\
   0.5 & 0.5 & 1    & Prolate \\
\hline                                   
\end{tabular}
\end{table}

\subsection{ Environmental dependence of the DM halo abundance function: Decontaminating the results from mass dependencies}\label{method-nomalisation}

\begin{figure}
\centering
\includegraphics[width=0.48\textwidth]{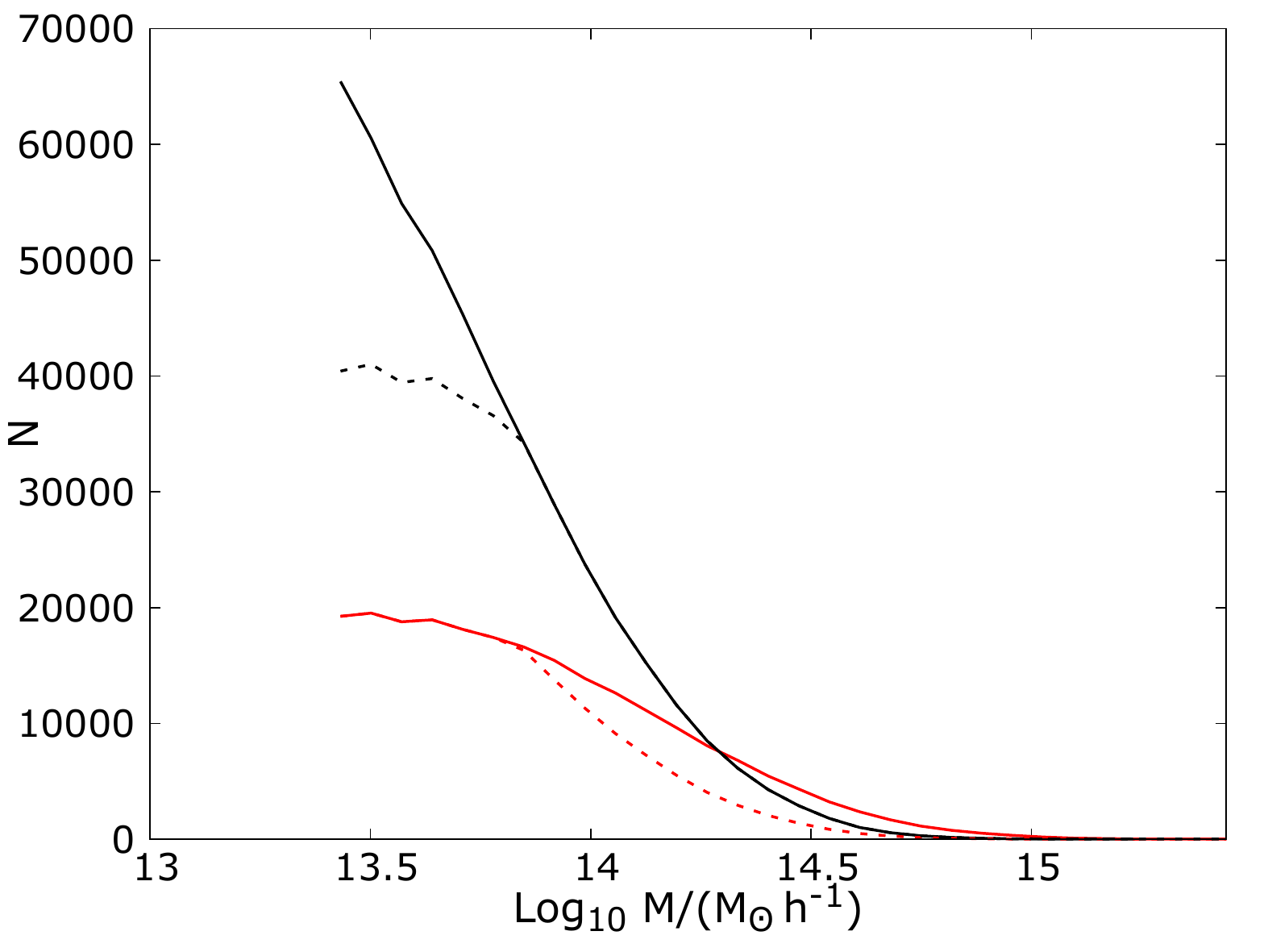}
\caption{Example of the original and the resulting AFs, after the matching procedure is applied, in order to obtain a mass-independent comparison of the dynamical parameters between two isolation bins, $1< R_{\rm isol}/r_{\rm vir}<4$ (red) and $4<R_{\rm isol}/r_{\rm vir}<7$ (black). On the vertical axis, $N$ is the number of halos. The solid lines correspond to the original AFs and the dashed lines to the AFs derived by the matching procedure. We note that a common shape of the AF between any two isolation bins is obtained and not the same overall number of halos.}
\label{mfs}
\end{figure}

In order to investigate how a specific property behaves in isolation, we separate our sample in bins of $R_{\rm isol}/r_{\rm vir}$ and we calculate the mean value and standard deviation of the parameter of interest at each isolation interval. However, the mass abundances of halos differ significantly for the different isolation bins (see details in \citet{Chira2018}) and, as expected, moving towards greater isolations means sampling a larger percentage of low-mass halos. Thus, in order to ensure that possible mass-dependencies of the dynamical parameters are factorized out from our environmental analysis, we used, at each isolation bin, a subsample of the halos that reproduces the shape of the mass function as in the other isolation bins. The procedure is based on a random sampling procedure, the result of which is a common abundance function (AFs) for both bins. To achieve statistically robust results, we re-sampled  the original abundance functions multiple times, using the jackknife procedure (Efron, 1979). An example of the normalization of the AFs is shown in Figure~\ref{mfs}.

To compare the parameters in all isolation states, we apply the above procedure of normalizing the AFs in a chain-wise fashion, that is, we start with AF$_1$-(1,4) and AF$_2$-(4,7) which result in the normalized abundance functions, AF$_1^{'}$ and AF$_2^{'}$. Subsequently, for the normalization of the AFs between the isolation intervals (4,7) and (7,10), instead of using AF$_2$ and AF$_3$, we used AF$_2^{'}$ and AF$_3$, so that the normalized AF$_2^{''}$ and AF$_3^{'}$ carry information from bins (1,4),(4,7), (7,10), and so on. Thus, moving towards larger isolations the normalized AFs "carry" information from all previous isolation bins.

Although the above chain-wise AF normalization does not provide a precise matching of the AF through the whole range of isolation bins, as it is not consistent backwards (e.g., isolation $(1,4)$ "carries" information only from $(4,7)$), we trust that it manages to decontaminate  our results from any halo mass dependencies as much as possible.

\subsection{Quantification of shape and spin alignments}\label{method-alignments}
A major part of the current work is focused on the study of the alignment between the spin and shape of halos as well as the respective vectors of neighboring halos. As a measure of such alignments we use the cosine of the misalignment angle formed by the vectors of interest in each case. The cosine is calculated as:
\begin{equation}
    \cos{\theta}= \frac{\Vec{v_1}\cdot\Vec{v_2}}{\abs{\Vec{v_1}}\abs{\Vec{v_2}}} \;,
\end{equation}
where $\Vec{v_1}$, $\Vec{v_2}$ are the vectors of interest in each case. In this work, we study three types of alignments: (a) shape - spin alignment between the principle axes and the spin-vector of a halo, (b) shape - shape alignment between the principle axes of neighboring halos, and (c) spin - spin alignment between the spin vectors of neighboring halos.

For the first and second case, the orientation and not the direction is relevant for the halo shape (defined by the principle axes of the fitted ellipsoidal). Thus, these cases of alignments are quantified using the absolute value of the cosine, $|\cos{\theta}|$, with $1$ corresponding to alignment and $0$ corresponding to perpendicular orientations. In the third case, it is useful to retain the sign since $1,0,-1$ correspond to parallel, perpendicular, and anti-parallel spin vectors, respectively.

To investigate the statistical tendencies regarding the discussed types of alignments, we carried out distributions of $\abs{\cos{\theta}}$ (or $\cos{\theta}$) and sought statistically significant deviations from a uniform distribution, which would be the case for random relative orientations. For the shape-spin and shape-shape alignments, an excess of the distribution towards values of $ \abs{\cos{\theta}} < 0.5 $ (i.e., $\theta < 45\degree$) corresponds to statistical preference for misalignment, while an excess of the distribution towards values of $\abs{\cos{\theta}} > 0.5 $ (i.e., $\theta > 45\degree$)  corresponds to statistical preference for alignment. For the case of spin - spin alignments, since the range of the misalignment angle is $0\degree< \theta < 180\degree$, an excess of the distribution towards values of $\cos{\theta} < 0 $ (i.e., $\theta > 90 \degree$) implies a statistical preference for anti-parallel spin vectors, while an excess of the distribution towards values of $\cos{\theta} > 0$ (i.e., $\theta < 90 \degree$) implies a statistical preference for parallel spin vectors. We note that for a clear comparison of the statistical tendencies between different halo subsamples, we show the normalized frequency distributions.

\subsection{Quantification of the statistical uncertainties}
We quantify the dependence of different halo properties on the environment, as well as on other physical properties, by estimating their mean values and its uncertainty in bins of the parameter of interest, according to the following:

Mean of parameter $Y,$ defined as:
\begin{equation}
   \langle Y \rangle = \frac{\sum_{i=1}^{N}Y_{i}}{N}\;,
\end{equation}
    \noindent
where N is the number of objects in each bin.

Precision of the estimated mean value will be provided by the standard error of the mean (SEM), defined as
\begin{equation}
{\rm SEM} = \frac{\sigma_{Y}}{\sqrt{N}}\;,
\end{equation}
where $\sigma_{Y}$ is the standard deviation of the data with respect to the estimated mean, defined as:
\begin{equation}
   \sigma_{Y} = \sqrt{\frac{\sum_{i=1}^{N}(Y_{i}-\langle Y \rangle)^2}{N-1}}.
\end{equation}

We note that when we present the frequency distributions, we use Poisson uncertainties, that is: $\sigma_{N} = \sqrt{N}$.

\section{Results}

\subsection{Dependence of DM halo properties on mass}

The dependence of halo dynamical properties on halo mass is a subject that is widely discussed in the literature. However, for reasons of completeness and consistency, we examine and discuss such dependencies in the context of our halos below.

For halo shape, we find that the mean flatness drops with halo mass, which means that higher mass halos tend to be more aspherical and more prolate than oblate, as indicated by the triaxiality parameter, a preference that becomes more prominent with increasing halo mass. These results are in agreement with the results of various relevant works in the literature \citeg{Bullock2001,Gottlober2006,AvilaReese2005, Shaw2006, Bailin2005, Allgood2006,Bett2007,VegaFerrero2017}.

\begin{figure}
    \centering
        \includegraphics[width=0.48\textwidth]{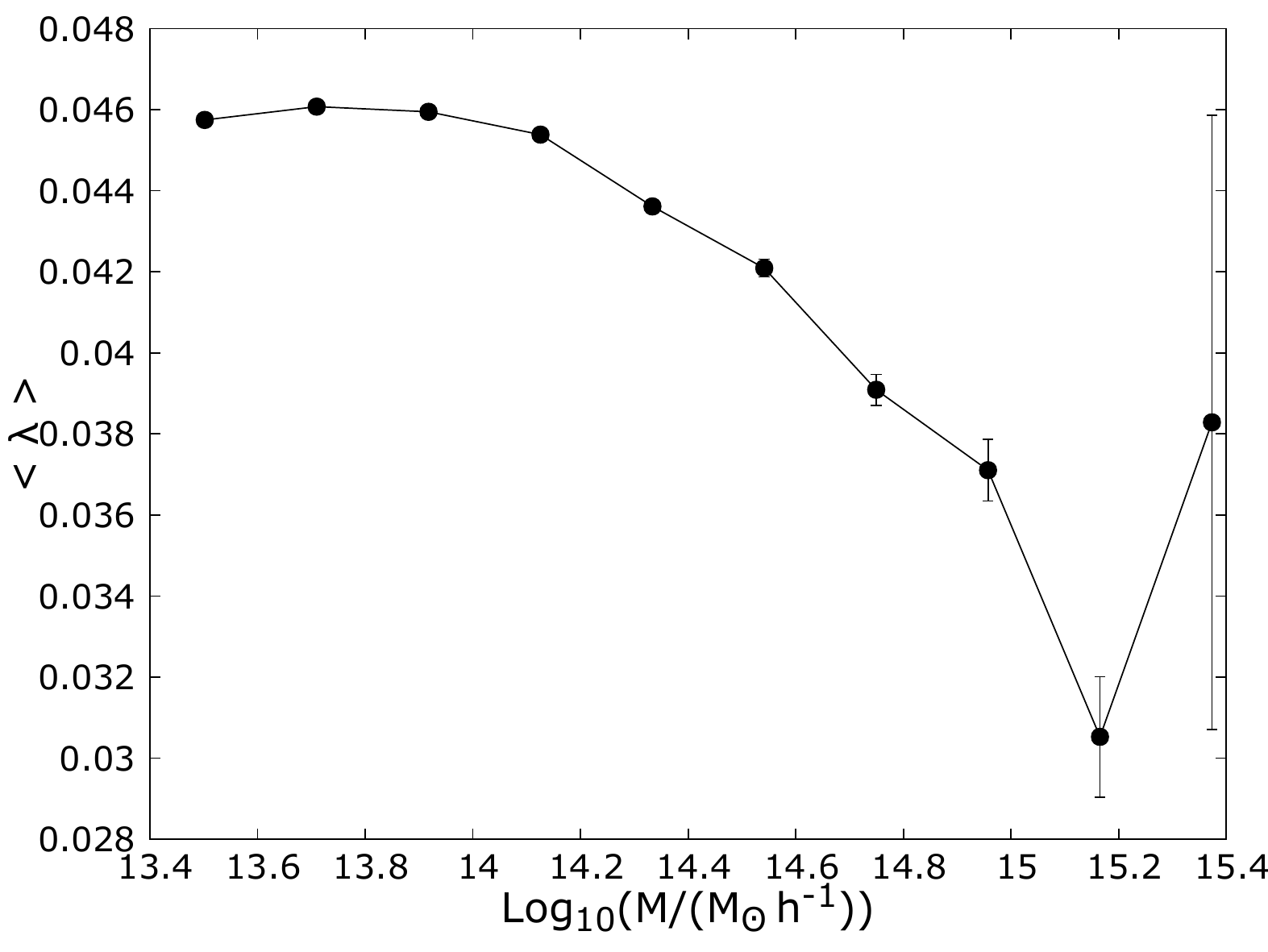}
    \caption{Dependence of halo spin parameter $\lambda$ on mass. The total sample of CHs is in ten mass bins of width of $\log{M/M_{\odot}}$. Here, we present the mean value of $\lambda$ in each mass bin with the errorbars corresponding to the $\pm {\rm SEM} $. We find that halo spin decreases systematically with halo mass.}\label{fig:spin-mass}
\end{figure}

For halo spin, we find that the spin of the prime axis decreases systematically with mass. Although previous authors \citeg{Knebe2008,Bett2007,Maccio2007,Munoz2011, Trowland2013} have only found such a dependency for high-redshift halos, we find a weak but systematic effect even for our relatively low-redshift ($z < 0.65$) DM halos, as can be clearly seen in Figure \ref{fig:spin-mass}.

Finally, for the virialization status, we wish to emphasize that the $|U/K|$ ratio should be considered only as a rough estimate of the virialization state of the DM halos (see discussion in Section \ref{method-properties}). However, even as a lower limit of the true value, the energy ratio should reveal tendencies among different mass halos. Indeed, $|U/K|$ is found to be closer to the virialization value, $|U/K| = 2$, for the lower- than for the higher-mass halos, which is in accordance with high-mass halos being dynamically more active than low-mass halos.

\subsection{Dependence of DM halo properties on halo isolation}
Here, we present our main results, which are related to the environmental dependence of the dynamical properties of our light-cone DM halos. In Figure \ref{fig:prop_isol}, we plot the mean value of each dynamical parameter of interest in bins of $R_{\rm isol}/r_{\rm vir}$. The indicated errorbars correspond to the standard error of the mean (SEM). Moreover, we present in the inset plots the frequency distribution of each property for the two extreme isolation bins, $1<R_{\rm isol}/r_{\rm vir}<4$ (red) and $30<R_{\rm isol}/r_{\rm vir}<70$ (blue). It is evident that although the two distributions overlap over a large range of parameter values, they are statistically distinct, as has also been confirmed by the Kolmogorov-Smirnov test.
We note that we use the SEM instead of standard deviation (which is much larger as evident from the width of the distribution in the inset plots) in order to highlight the systematic and statistically important change of the mean value in the different isolation bins, which corresponds to a small but systematic change in the distributions with isolation.
\subsubsection{Shape}

\noindent
With regard to the flatness, $c/a$,
we investigate the ratio of the minor over the major axis of the fitted ellipsoid of the DM halos for different isolation states described in Section \ref{environment}.
As shown in Figure~\ref{fig:c_a}, we find a systematically decreasing tendency of the mean value $\langle c/a\rangle$ with isolation, implying that isolated halos are statistically flatter than those residing in dense regions.
As for the triaxiality parameter, $T$,
as shown in Figure~\ref{T_isol} we find that the triaxiality parameter has, for all isolation bins, a mean value of $0.7\lesssim \langle T\rangle \lesssim 0.8$, meaning that in all isolation states halos are more prolate than oblate. The more isolated the halos, the more prolate they are.

These results indicate that halo shapes are affected by the environment and are shown to be more aspherical and more prolate in lower density environments, which is in agreement with the conclusions of \citet{Bett2007}. The deviation from sphericity and the preference for prolateness of DM halos can be attributed to anisotropic collapse and mass infall along specific directions, those defined by the filamentary large-scale structure where halos are initially formed. The anisotropic structure of the local environment is expected to survive more efficiently in "quiet" environments (high-isolation regions). On the other hand, low-isolation halos undergo more violent interactions, which tend to sphericalize their shape.

\subsubsection{Spin of the prime axis}
The ependence of halo spin on environment is an area that sees conflicting results. As shown in Figure~\ref{spin}, the mean spin of the DM halos drops systematically with increasing isolation radius. Such an effect seems intuitive since the strong interactions and merging effects that occur in high-density (low-isolation) environments could be responsible for the acquisition of angular momentum. Our results are qualitatively in agreement with those of \citet{Bett2007, Wang2011, Lee2017, Johnson2019}. However, other authors have reported different results; \citet{AvilaReese2005} find that their {\em CLUSTER} halos have lower spin values than their {\em VOID} halos, while \citet{Maccio2007} find no dependence of halo spin on environment.

In these aforementioned studies, authors have used a variety of different approaches for the quantification of halo clustering and of halo environment which could be, at least partially, a source for the discrepancies of their results, as has been shown by \citet{Libeskind2018}. Moreover, the dependence of halo-spin on the environment has been found to correlate also with halo mass, namely, it is stronger for massive halos \citeg{Wang2011}, which possibly explains the results of \citet{Maccio2007} based on relatively low-mass halos.

\subsubsection{Virialization via $U/K$}
As a measure of the virialization status we compute the ratio, $|U/K|$, of the potential over the kinetic term of the total energy of our halos which, according to the virial theorem, should be $|U/K| \approx 2$ for fully virialized halos.

\begin{figure*}
    \centering
    \begin{subfigure}[b]{0.48\textwidth}
    \includegraphics[width=\textwidth]{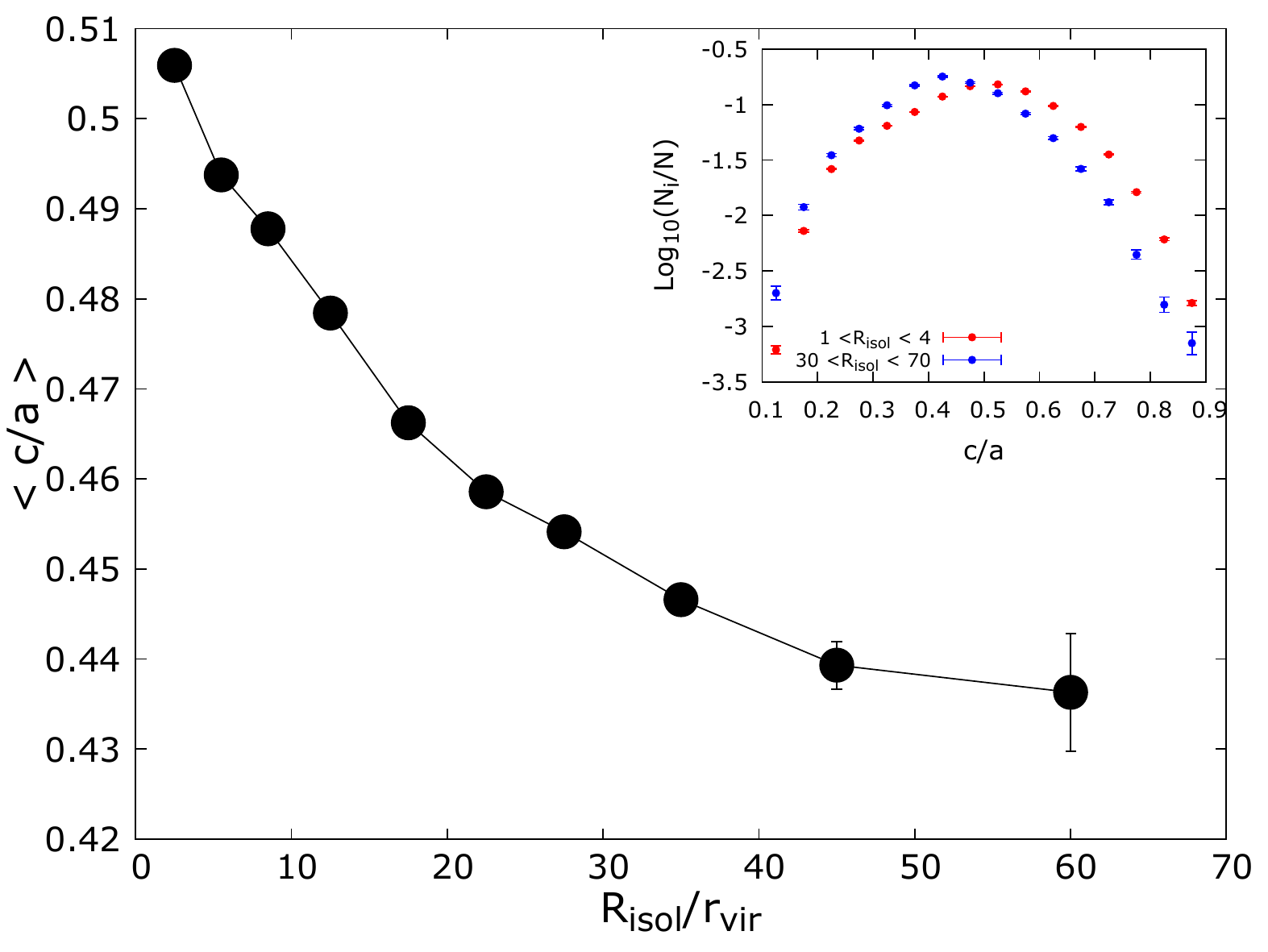}
    \caption{Mean flatness, $\langle c/a\rangle$, with isolation: We find a systematic decrease with increasing isolation radius, implying that more isolated objects are statistically less spherical than those residing in dense regions.}
    \label{fig:c_a}
    \end{subfigure}
    ~ 
    \begin{subfigure}[b]{0.48\textwidth}
        \includegraphics[width=\textwidth]{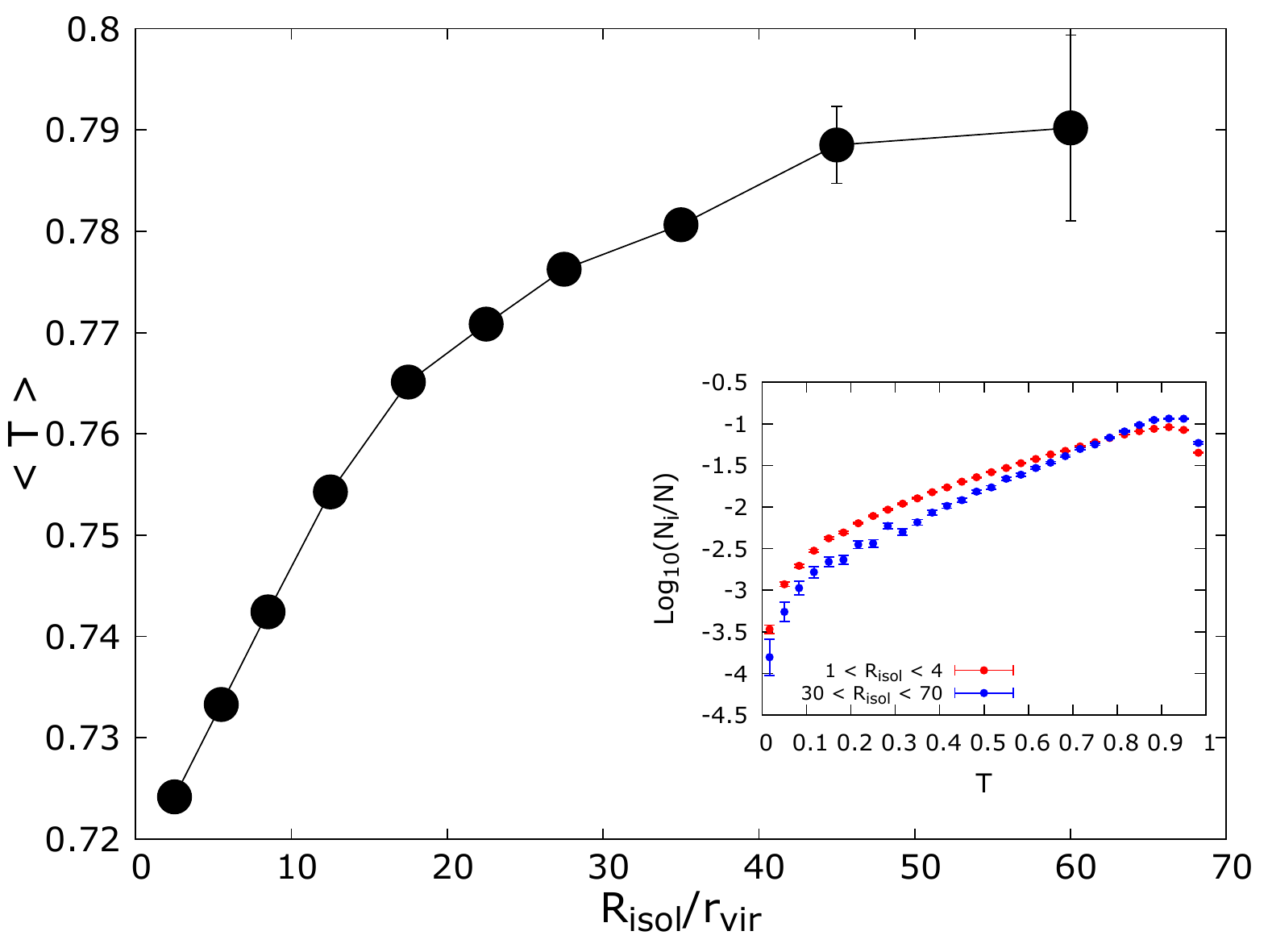}
        \caption{Mean Triaxiality, $\langle T\rangle$, with isolation: A systematic increase of $\langle T\rangle$ with increasing isolation radius is observed, implying that isolated objects are statistically more prolate than those residing in dense regions.}
    \label{T_isol}
    \end{subfigure}
    ~ 
    \begin{subfigure}[b]{0.48\textwidth}
        \includegraphics[width=\textwidth]{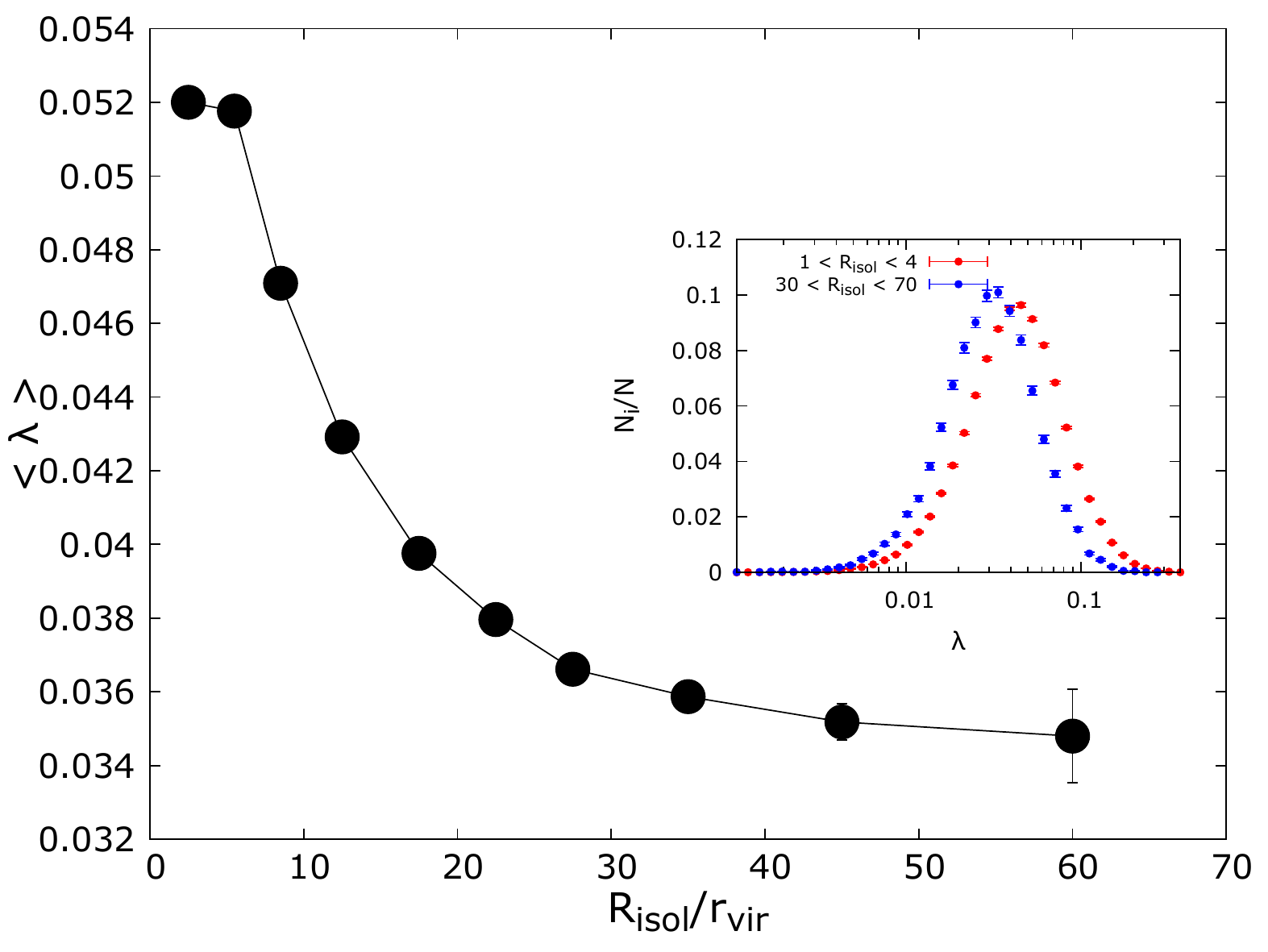}
        \caption{Mean spin, $\langle \lambda \rangle$, with isolation: A systematic decrease of $\langle \lambda\rangle$ with increasing halo isolation is observed.}
    \label{spin}
    \end{subfigure}
    \begin{subfigure}[b]{0.48\textwidth}
        \includegraphics[width=\textwidth]{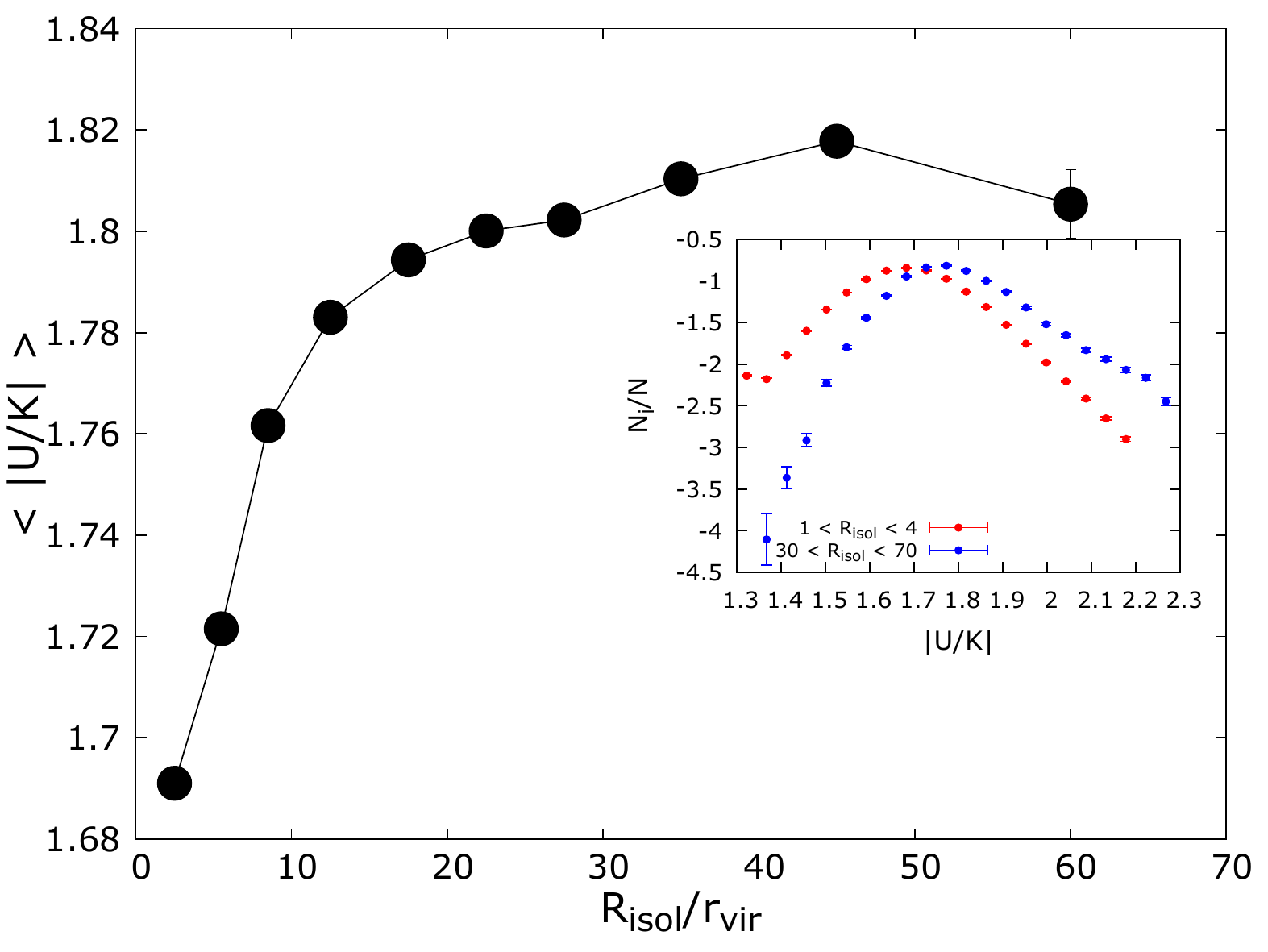}
        \caption{Virialization status, $\langle |U/K|\rangle$, with isolation: A systematic increase of $\langle |U/K|\rangle$ with halo isolation is observed. The increase is significantly faster for the lower isolation bins ($R_{\rm isol}/r_{\rm vir}\lesssim 15$).}
        \label{fig:vir}
    \end{subfigure}
    \caption{Dependence of DM halo dynamical properties on halo isolation. The frequency distributions of each property, for the two extreme isolation bins $1<R_{\rm isol}/r_{\rm vir}<4$ (red) and $30<R_{\rm isol}/r_{\rm vir}<70$ (blue), are presented in the inset plots.} \label{fig:prop_isol}
\end{figure*}

As shown in Figure~\ref{fig:vir}, we find a significant dependence for $\langle|U/K|\rangle$ on isolation, with the lower isolation bins corresponding to lower values of $|U/K|$, that is, the smaller the isolation radius, the higher the deviation from virialization. This is probably a consequence of the intense tidal fields and strong interactions in the high-density regions, which become quieter with increasing halo-isolation. For isolation values higher than $R_{\rm isol}/r_{\rm vir} \gtrsim 20$, halos are significantly isolated and their virialization status does not seem to be strongly affected by isolation.

\subsection{Correlations between DM halo properties}

\begin{figure*}
    \centering
    \begin{subfigure}[b]{0.45\textwidth}
        \includegraphics[width=\textwidth]{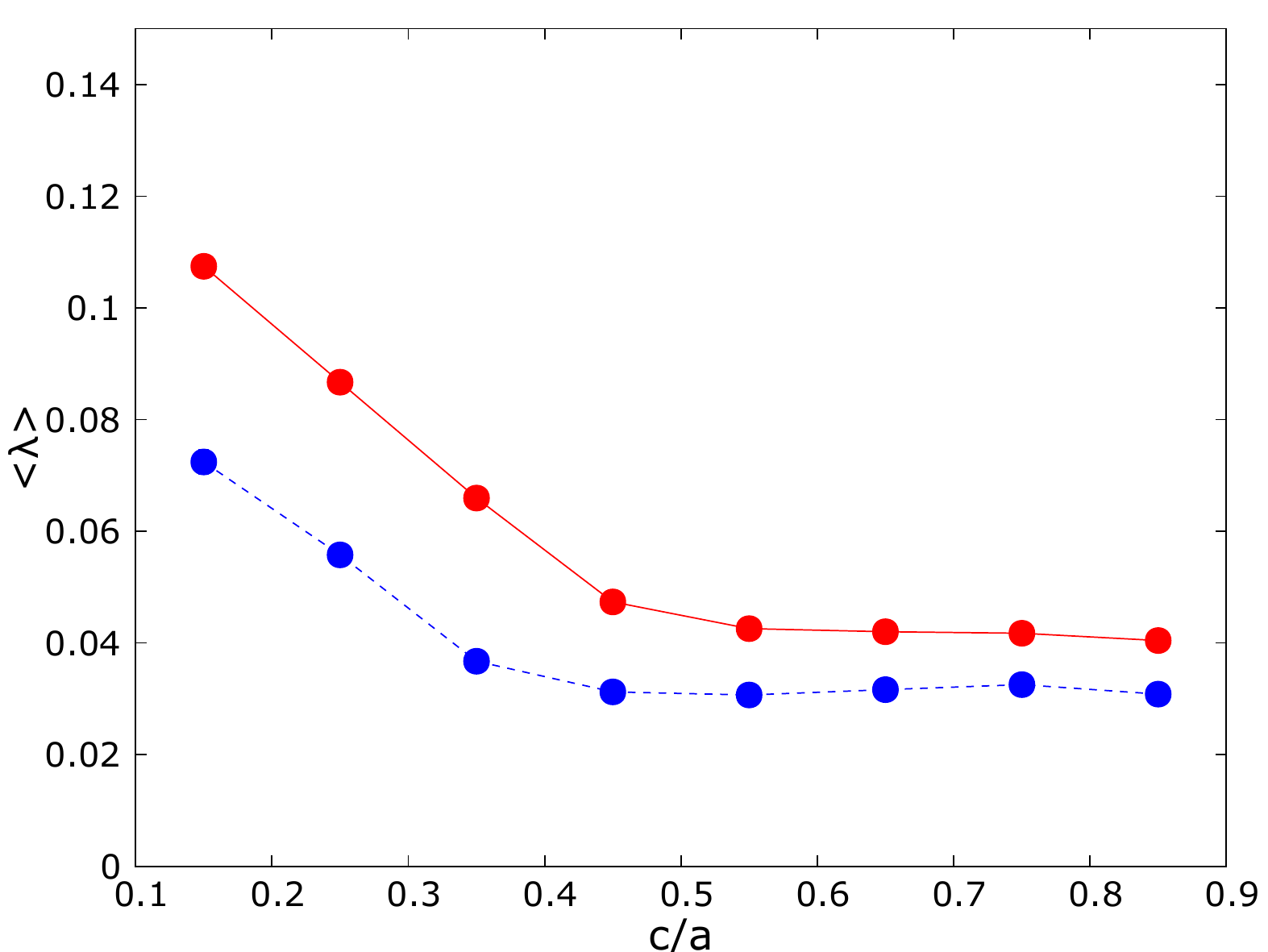}
        \caption{\emph Flatness vs spin}
        \label{fig:flat_spin}
    \end{subfigure}
    ~ 
    \begin{subfigure}[b]{0.45\textwidth}
      \includegraphics[width=\textwidth]{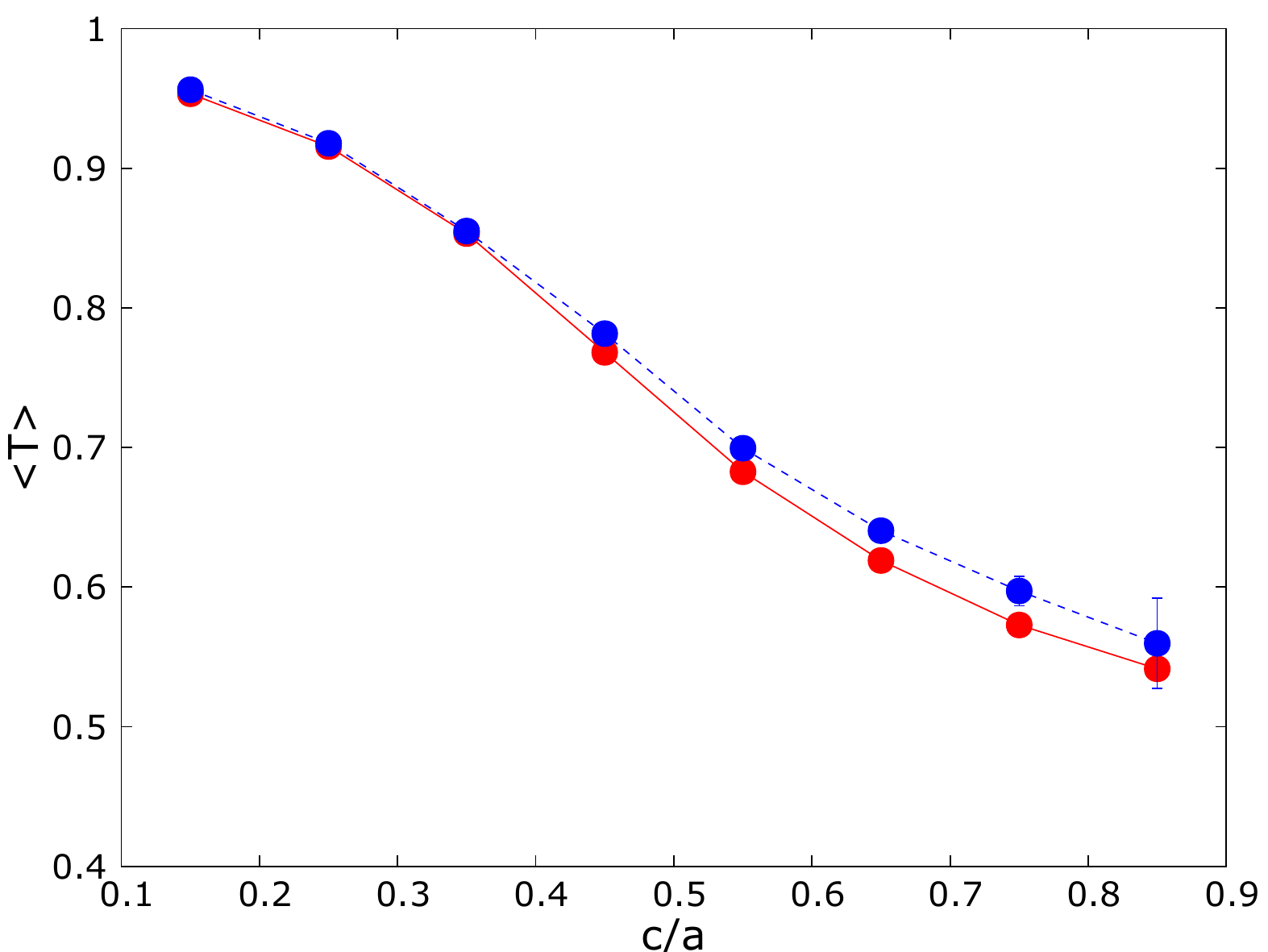}
        \caption{Flatness vs triaxiality}
        \label{fig:flat_T}
    \end{subfigure}
    ~ 
    \begin{subfigure}[b]{0.45\textwidth}
        \includegraphics[width=\textwidth]{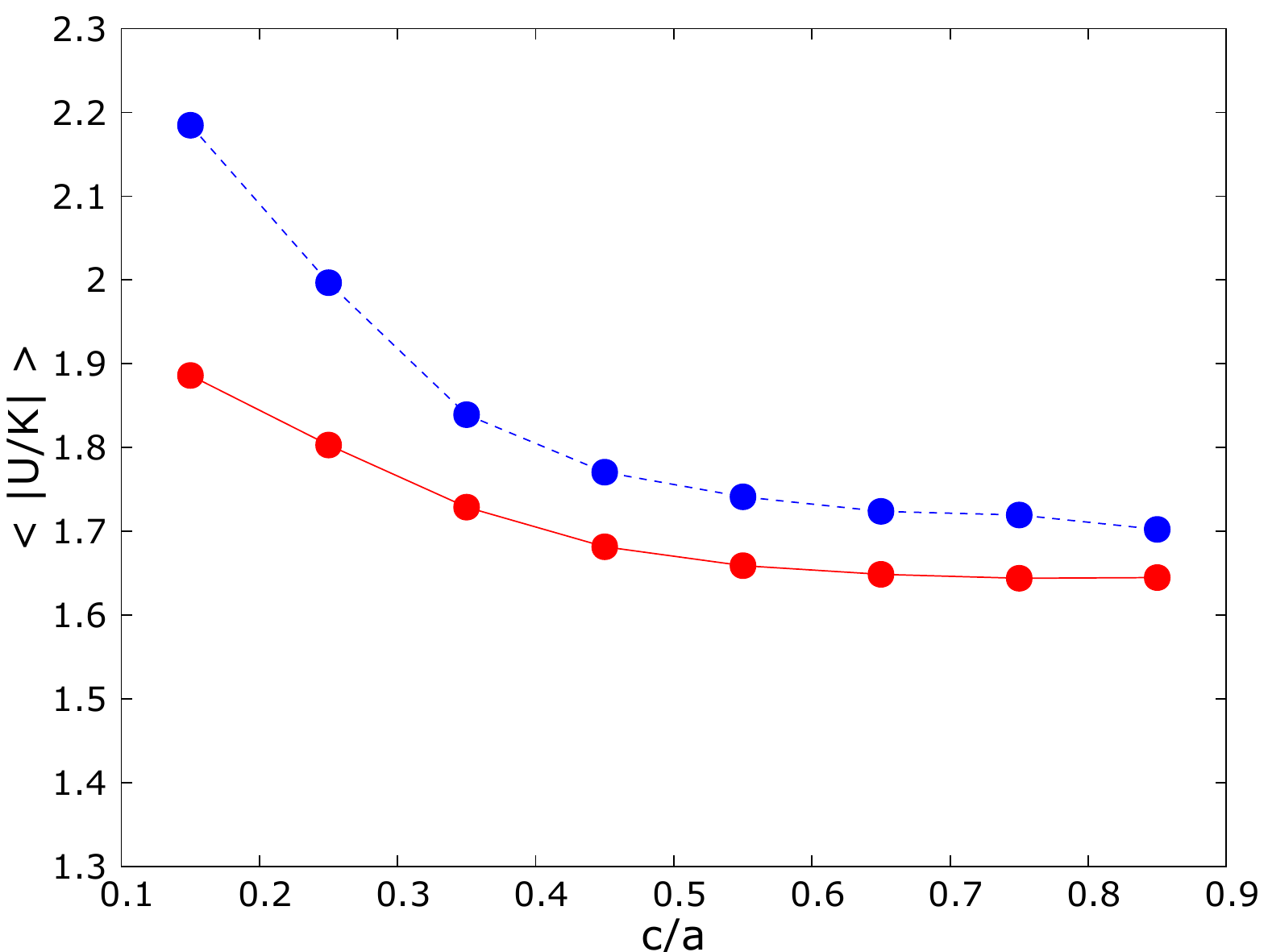}
        \caption{Flatness vs energy ratio}
        \label{fig:flat_vir}
    \end{subfigure}
    \begin{subfigure}[b]{0.45\textwidth}
        \includegraphics[width=\textwidth]{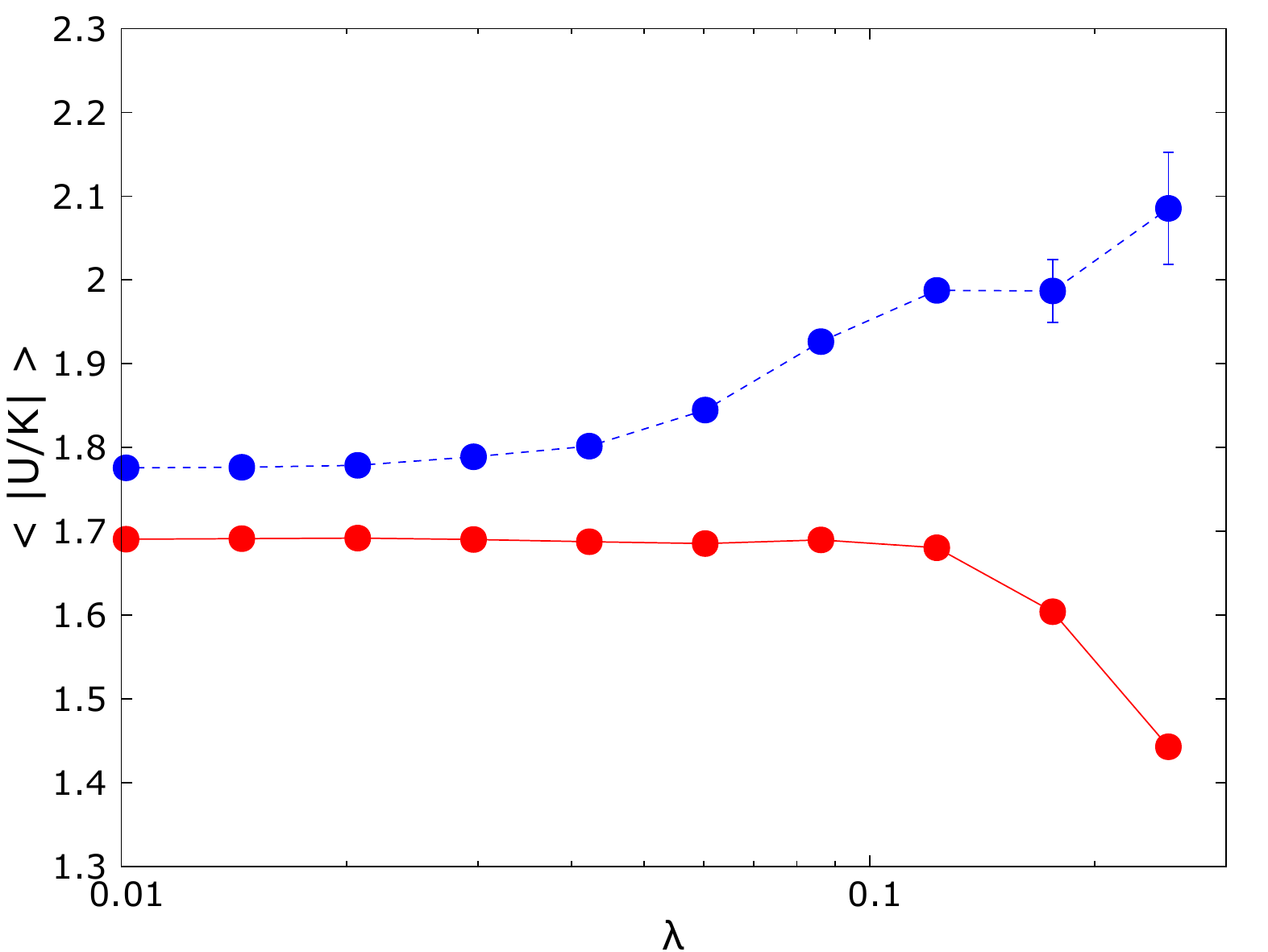}
        \caption{Spin vs energy ratio}
        \label{fig:spin_vir}
    \end{subfigure}
    \begin{subfigure}[b]{0.45\textwidth}
        \includegraphics[width=\textwidth]{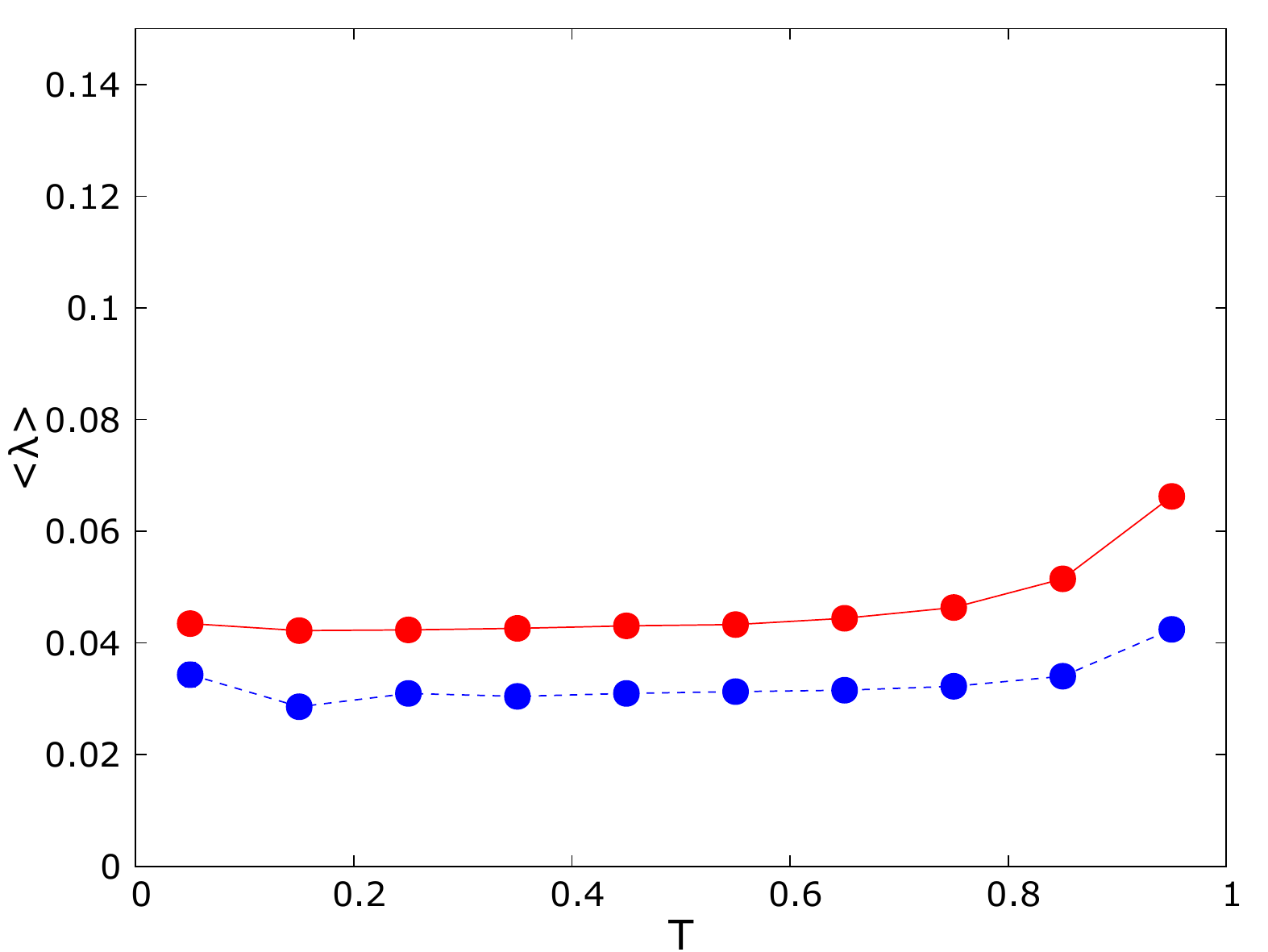}
        \caption{Triaxiality vs spin}
        \label{fig:spin_T}
    \end{subfigure}
    \begin{subfigure}[b]{0.45\textwidth}
        \includegraphics[width=\textwidth]{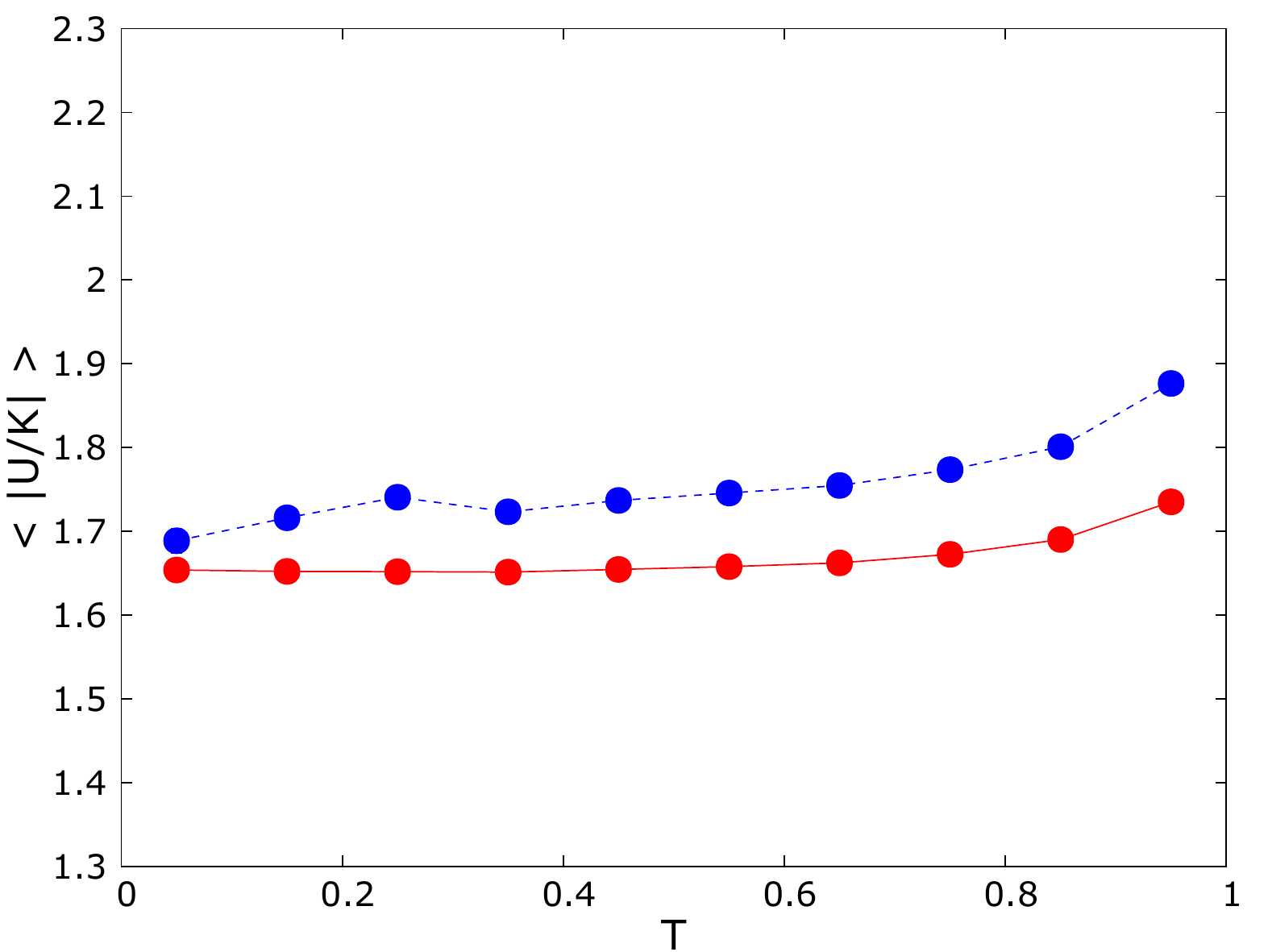}
        \caption{Triaxiality vs energy ratio}
        \label{fig:T_vir}
    \end{subfigure}
    \caption{Correlations of DM halo properties. Red (solid line) corresponds to the lowest isolation bin, $1<R_{\rm isol}/r_{\rm vir}<4$ and blue (dashed line) corresponds to the highest isolation bin, $30<R_{\rm isol}/r_{\rm vir}<70$.}\label{fig:prop_depend}
\end{figure*}

In this part of the paper, we investigate the correlation between the dynamical properties of our DM halos and their possible dependence on the environment. In Figures ~\ref{fig:flat_spin}-\ref{fig:T_vir}, on the vertical axis, we present the mean value of the property of interest in bins of the property on the horizontal axis and the error bars correspond to the $\rm SEM$. The different colors denote the two extreme isolation bins, $1<R_{\rm isol}/r_{\rm vir}<4$ -red, and $30<R_{\rm isol}/r_{\rm vir}<70$ -blue. 

We find that the spin parameter of the prime axis decreases systematically with flatness (Fig. \ref{fig:flat_spin}), meaning that the rounder the DM halos are, the slower they spin. This dependence of the spin parameter on flatness is independent of the halo isolation. These results are in qualitative agreement with \citet{Bett2007, Johnson2019}. Moreover, the triaxiality parameter decreases systematically with flatness (Fig. \ref{fig:flat_T}). The anti-correlation of the two parameters is expected by their definition.

Regarding the virialization state of DM halos, expressed by $\langle|U/K|\rangle$, it is inversely correlated with flatness (Fig. \ref{fig:flat_vir}). This is evident in both environments, although highly isolated halos are more virialized, indicated by the higher on average value of $U/K$. The value of $|U/K|$ decreases from the theoretical virial value of $|U/K| \simeq 2$ as halos become more spherical. Although this seems counter-intuitive, it could well be explained by the fact that highly elongated clusters $(c/a\lesssim 0.25)$ reflect an unvirialized stage, being at maximum separation of the merging components, in which the potential energy is relatively higher than average; therefore, the $U/K$ is higher than the corresponding value for less elongated clusters.

In Fig, \ref{fig:spin_vir}, for $0.01<\lambda<0.1,$ we find no correlation between spin and virialization status. However, for high-spin values ($\lambda>0.1$), we find a distinct behavior of the low and high-isolation halos: a systematic decrease of $\langle |U/K|\rangle$ with increasing spin for the low-isolation halos (red points), which could be attributed to strong interactions and mergers, while for the isolated halos (blue points), the high spin halos have higher $|U/K|$, consistent with the nominal virialization value ($\simeq 2$), and possibly suggesting that in the absence of interactions and mergers, a relatively high value of $\lambda$ corresponds to virialized halos. In addition, we find that the spin and triaxiality parameters (Fig. \ref{fig:spin_T}) do not show a significant correlation except for the highest-$T$ halos which have higher spins, possibly pointing to the origin of high prolateness being due to major mergers. Finally, there is no significant trend between $\langle |U/K|\rangle$ and $T$ (Fig. \ref{fig:T_vir}).

We note that in Figures \ref{fig:flat_vir}-\ref{fig:spin_vir}, in the case of isolated halos (blue) for the lower-$c/a$ and the higher-$\lambda$ bins, the computed values of $\langle |U/K|\rangle$ exceed the virialization limit of $2$, implying that the corresponding halos are unbound. We wish to emphasize that we approach the virialization parameter with reservation since the way of calculating the potential energy of DM halos is inevitably a rough approximation, as discussed previously.

\subsection{Spin-shape alignment}

\begin{figure}
    \centering
    \includegraphics[width=0.48\textwidth]{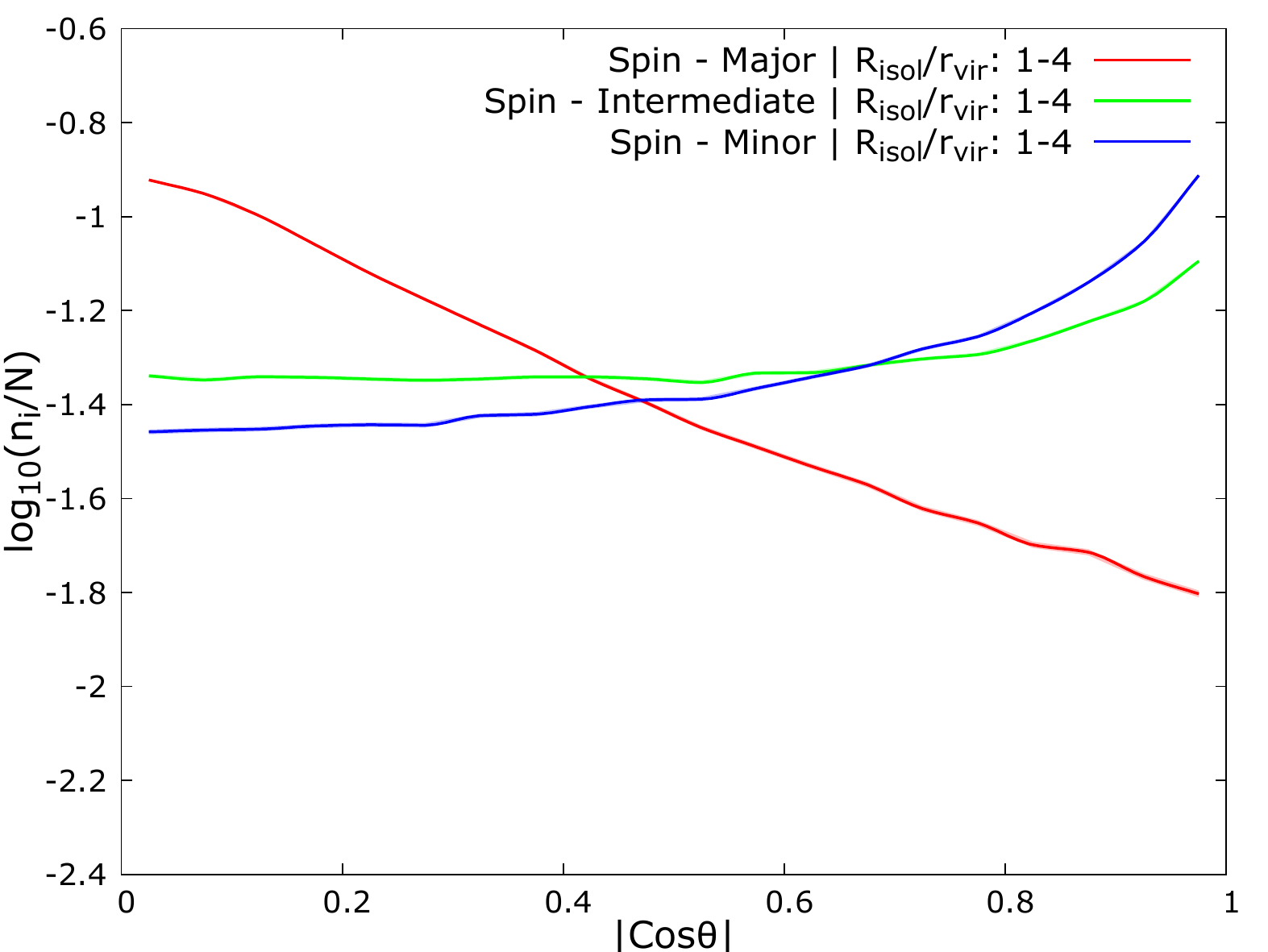}\vfill
    \includegraphics[width=0.48\textwidth]{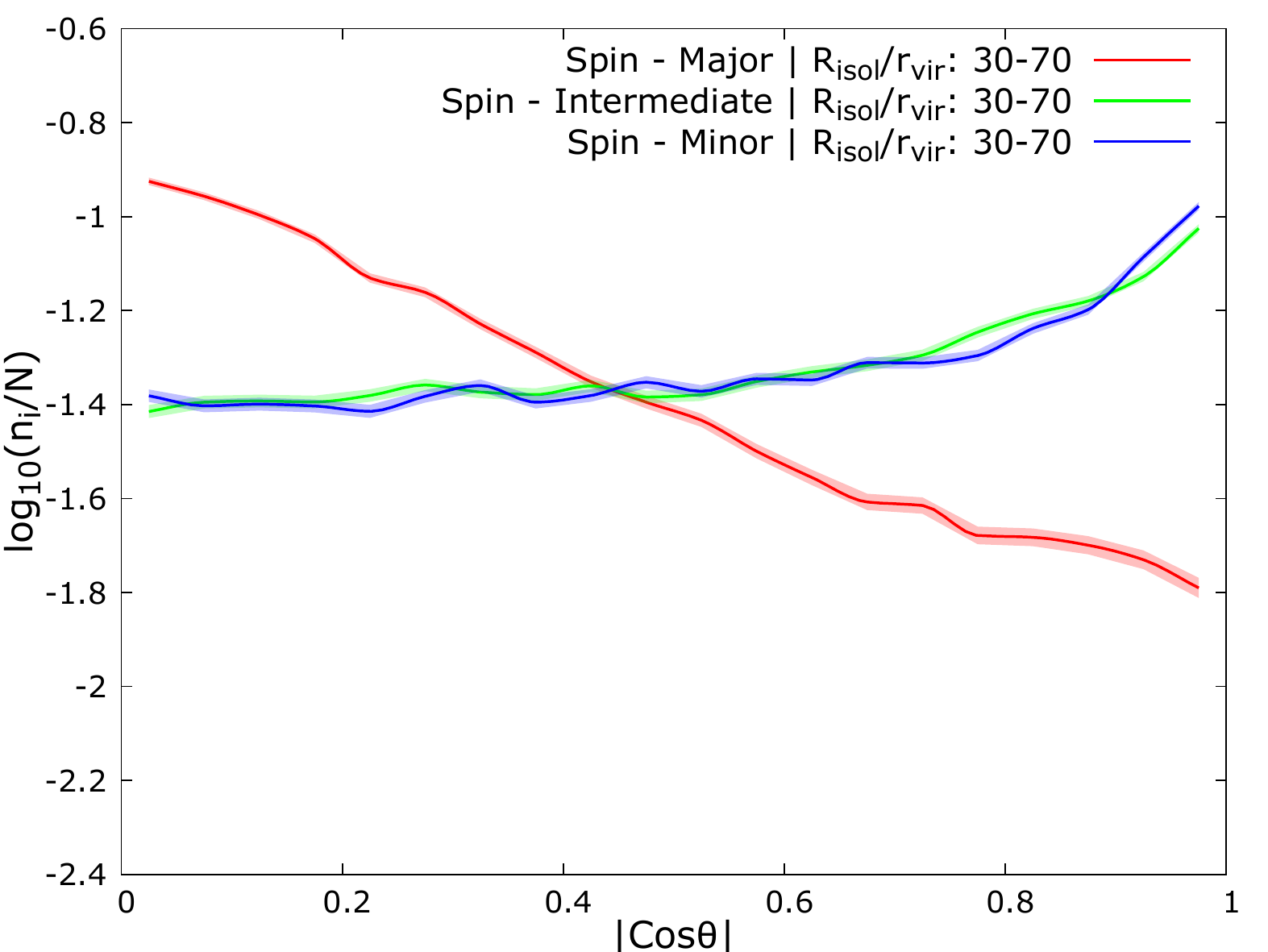}
    \caption{ Frequency distribution of $|\cos{\theta}|$, where $\theta$ is the angle between the spin vector and the major (red), intermediate (green), minor (blue) axis of the DM halos, for two distinct halo environments: {\it Upper panel:} Halos in dense environment ($1<R_{isol}/r_{vir}<4$). {\it Lower panel:} Extremely isolated halos ($30<R_{isol}/r_{vir}<70$). The shaded splines denote the Poisson uncertainties.}
    \label{fig: cos_distr_isol}
\end{figure}

In this section, we explore the alignment of the principal axes of DM halos with their spin vector and consider how this alignment varies with halo isolation. In Figure \ref{fig: cos_distr_isol}, we present the frequency distribution of the cosine of the misalignment angle between the halo axes and the spin vector for halos with isolation $1<R_{\rm isol}/r_{\rm vir}<4$ and $30<R_{\rm isol}/r_{\rm vir}<70$. The width of the shaded splines denotes the Poisson uncertainty of the distributions. We find that for halos in both isolation bins the spin vector tends to be perpendicular to the halo major axis. In the case of low-isolation halos (upper panel), the spin vector is more likely to be more aligned with the halo minor axis, although some alignment is also evident with the intermediate axis -- an observation that is more pronounced for the extremely isolated halos (lower panel). This can be explained by the fact that prolateness increases with isolation (see Figure \ref{T_isol}) and, thus, the minor and intermediate axes are statistically equivalent for the extremely isolated halos.

We further repeat the above exercise for two halo populations sampled according to the shape of the CHs, as defined using the triaxiality parameter:
\begin{itemize}
    \item {Population A:} Prolate halos ($T > 0.8$)
    \item {Poplulation B:} Oblate halos ($T < 0.25 $)
\end{itemize}
We note that our prolate criterion $(T>0.8)$ is stricter (compared to the oblate criterion, $T<0.25$) since their large number in our catalog allows us to be more selective and still maintain a sufficient number of halos for statistical purposes. Throughout the rest of this paper, we use the $T$-ranges mentioned above in the definition of Populations A and B to classify CHs (and their neighbors) as prolate or oblate.

\begin{figure}
    \centering
    \includegraphics[width=0.48\textwidth]{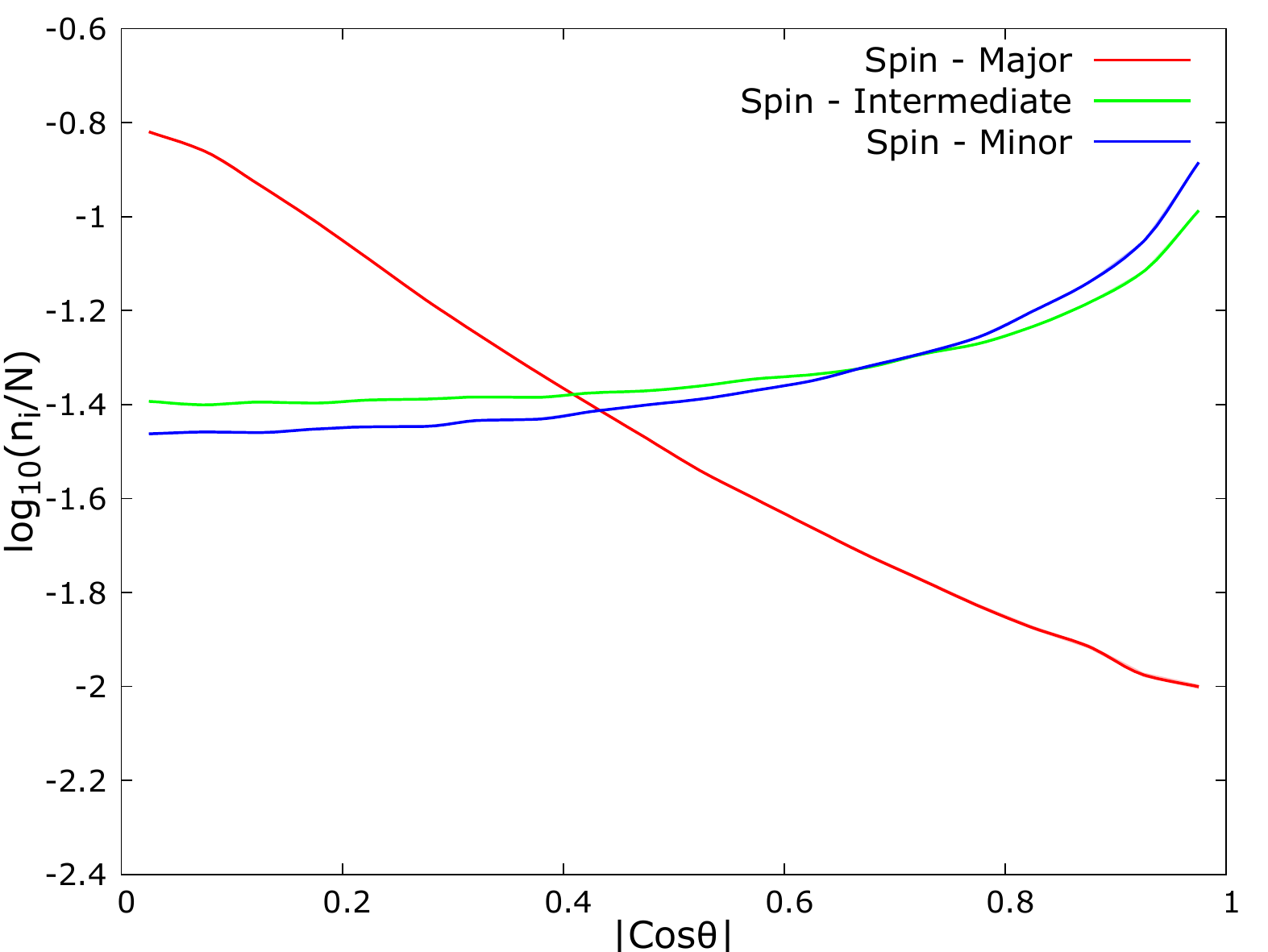}\vfill
    \includegraphics[width=0.48\textwidth]{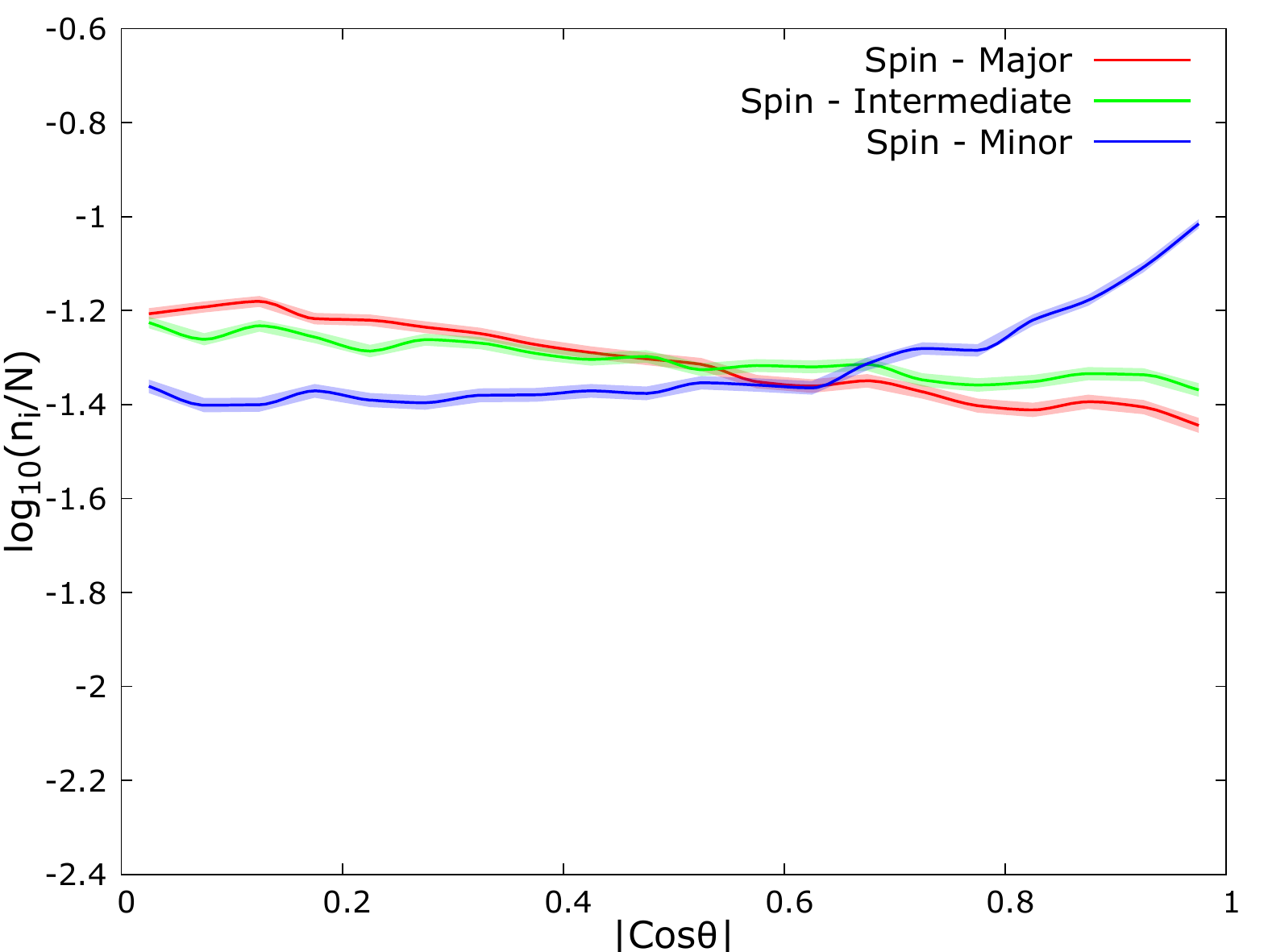}
    \caption{Frequency distribution of $|\cos{\theta}|$, where $\theta$ is the angle between the spin vector and the major (red), intermediate (green), minor (blue) axis of the DM halos for two distinct halo populations based on their shape: {\it Upper panel:} Population A (prolate halos, $T>0.8$). {\it Lower panel:} Population B (oblate halos, $T<0.25)$.}
    \label{Cos_distr_shapes}
\end{figure}

In Fig. \ref{Cos_distr_shapes}, we present the distribution of $|\cos{\theta}|$ separately for populations A and B to find that prolate halos (upper panel) are equally likely to have their spin aligned with the minor and intermediate axis, that is, having a "boomerang-like" rotation. Oblate halos (lower panel) have their spin vectors aligned predominantly with their minor axis and perpendicular to both the intermediate and major axes, that is, they show a "spinning-top-like" rotation.

\subsection{Shape-shape alignment}

\begin{figure}
\centering
\includegraphics[width=0.48\textwidth]{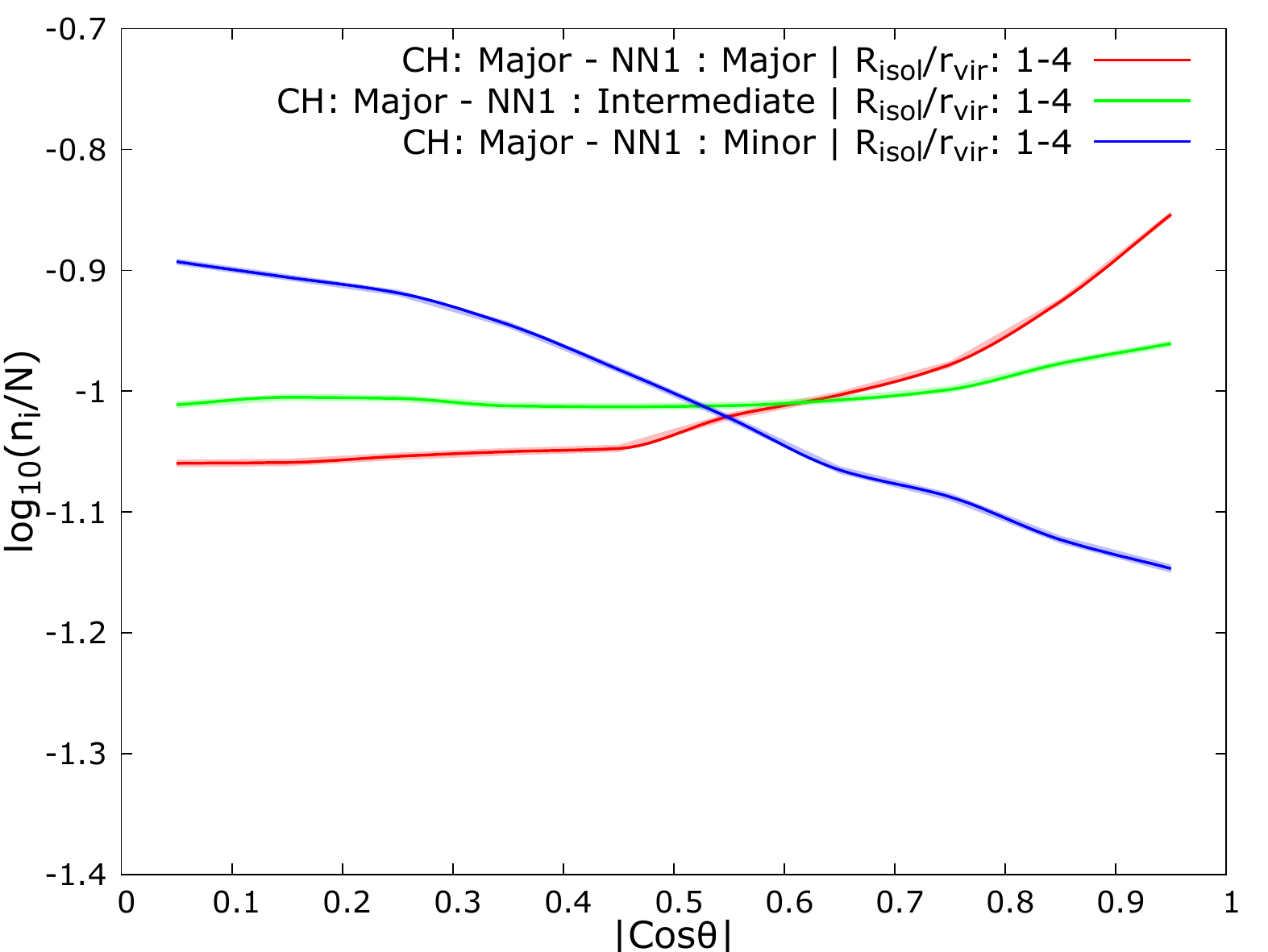}\hfill\\
\includegraphics[width=0.48\textwidth]{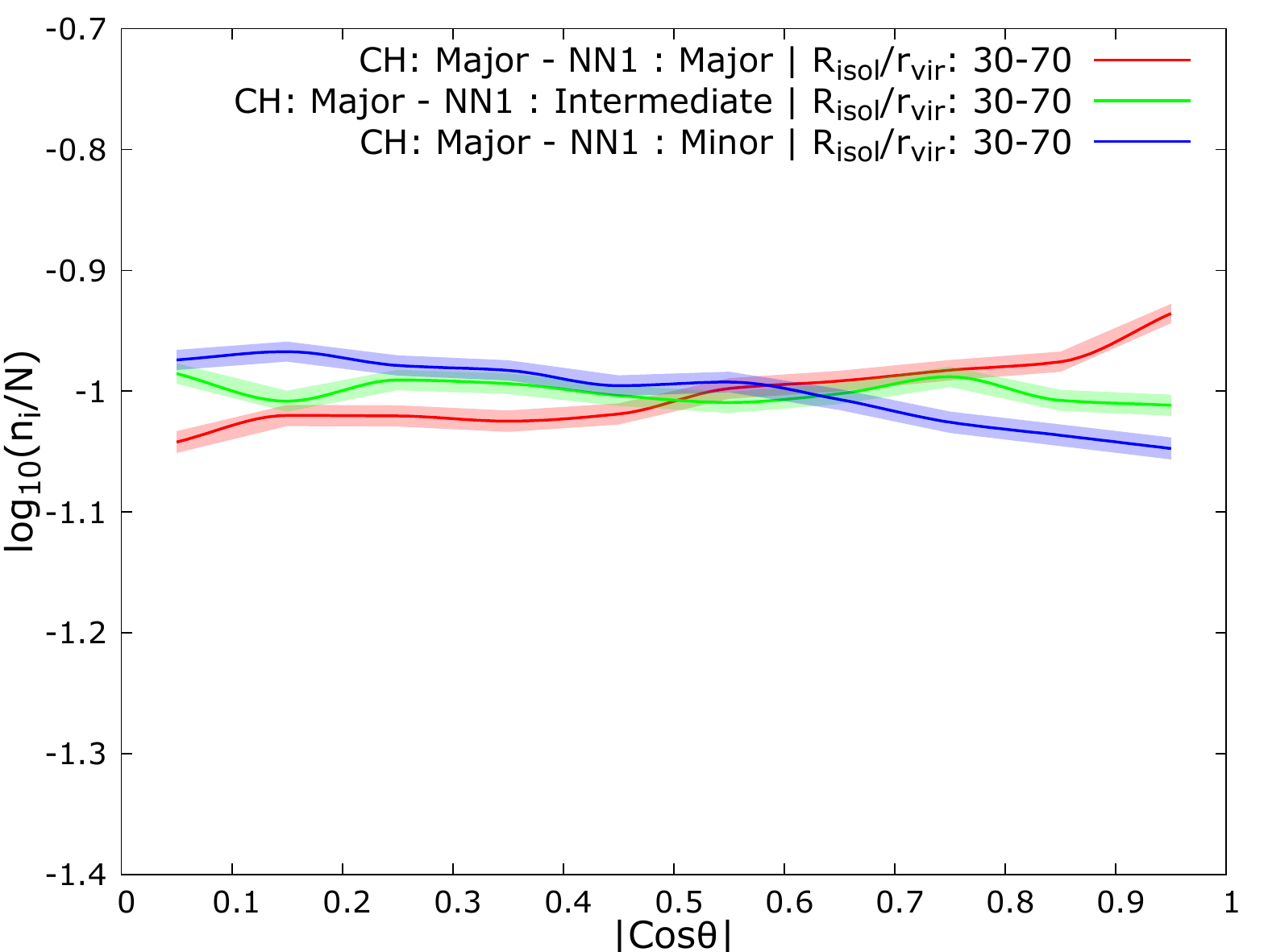}
\caption{Frequency distribution of the $|\cos(\theta)|$, where $\theta$ is the angle between the major axis of the CH and the major (red), intermediate (green), and minor (blue) axis of its first nearest neighbor (NN1). We show two extreme isolation bins: close halo pairs where the CHs are separated from their NN1 by $1<R_{\rm isol}/r_{\rm vir}<4   $ ({\em upper panel}), and extremely isolated halos with CHs separated from their NN1 by $30<R_{\rm isol}/r_{\rm vir}<70$ ({\em lower panel}).}
\label{maj-nn}
\end{figure}

In this section, we investigate the axial alignment of the closest halo neighbors and its dependence on halo isolation status. In Figure \ref{maj-nn}, we present the frequency distributions of $|\cos{\theta}|$, where $\theta$ is the misalignment angle between the major axis of the CH and the principal axes (major-red, intermediate-green, and minor-blue) of its first nearest neighbor (NN1). For the case of close halo pairs ($1<R_{\rm isol}/r_{\rm vir}<4$, upper panel), we find a strong statistical tendency for alignment of the major axes of the DM halo closest neighbors. This tendency for alignment weakens with increasing isolation ( $30<R_{\rm isol}/r_{\rm vir}<70$, lower panel). For the latter case of extremely isolated halos, halo separations correspond to crossing times longer than the Hubble time (spanning from $t_{H_{z=0.65}}=7.591 Gyr$ to $t_{H_{0}}=13.517 Gyr$ for the assumed cosmology), hence, the three distributions are almost identical. However, a slight excess corresponding to major axes alignment is still evident, indicating that even at such large distances the nearest halo neighbors retain some memory of the common origin from the initial anisotropic matter distribution within which they were formed.

We also study the alignment properties of close halo pairs ($1<R_{\rm isol}/r_{\rm vir}<4$) as a function of the pair isolation status, since we have verified that they can be found in a wide range of isolation states. The isolation criterion of the pair that we use is the normalized distance of the CH to its second nearest neighbor (NN2), that is, $R_{\rm isol_{2}}/r_{\rm vir}$.

\begin{figure}
\centering
\includegraphics[width=0.48\textwidth]{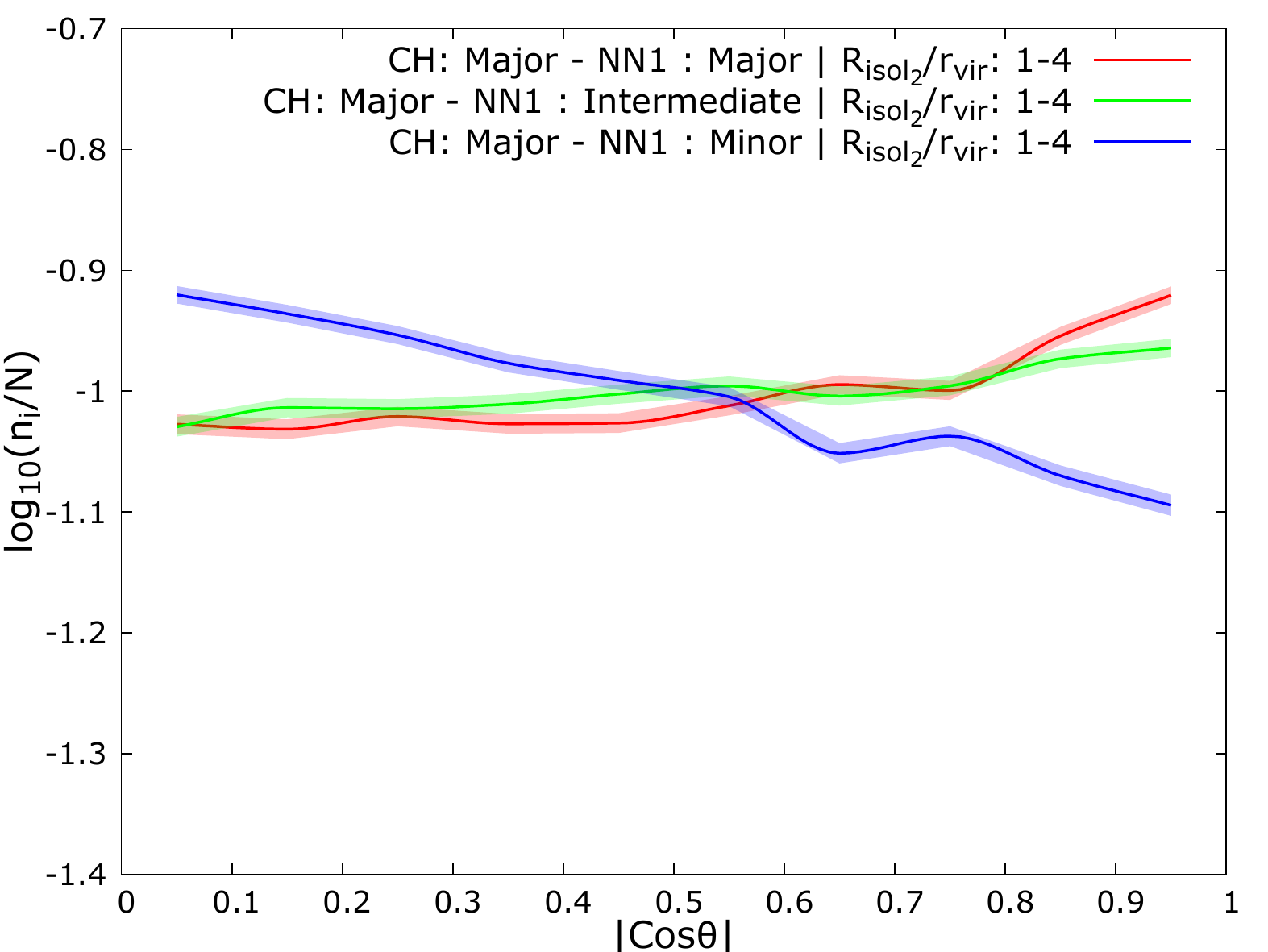}\hfill\\
\includegraphics[width=0.48\textwidth]{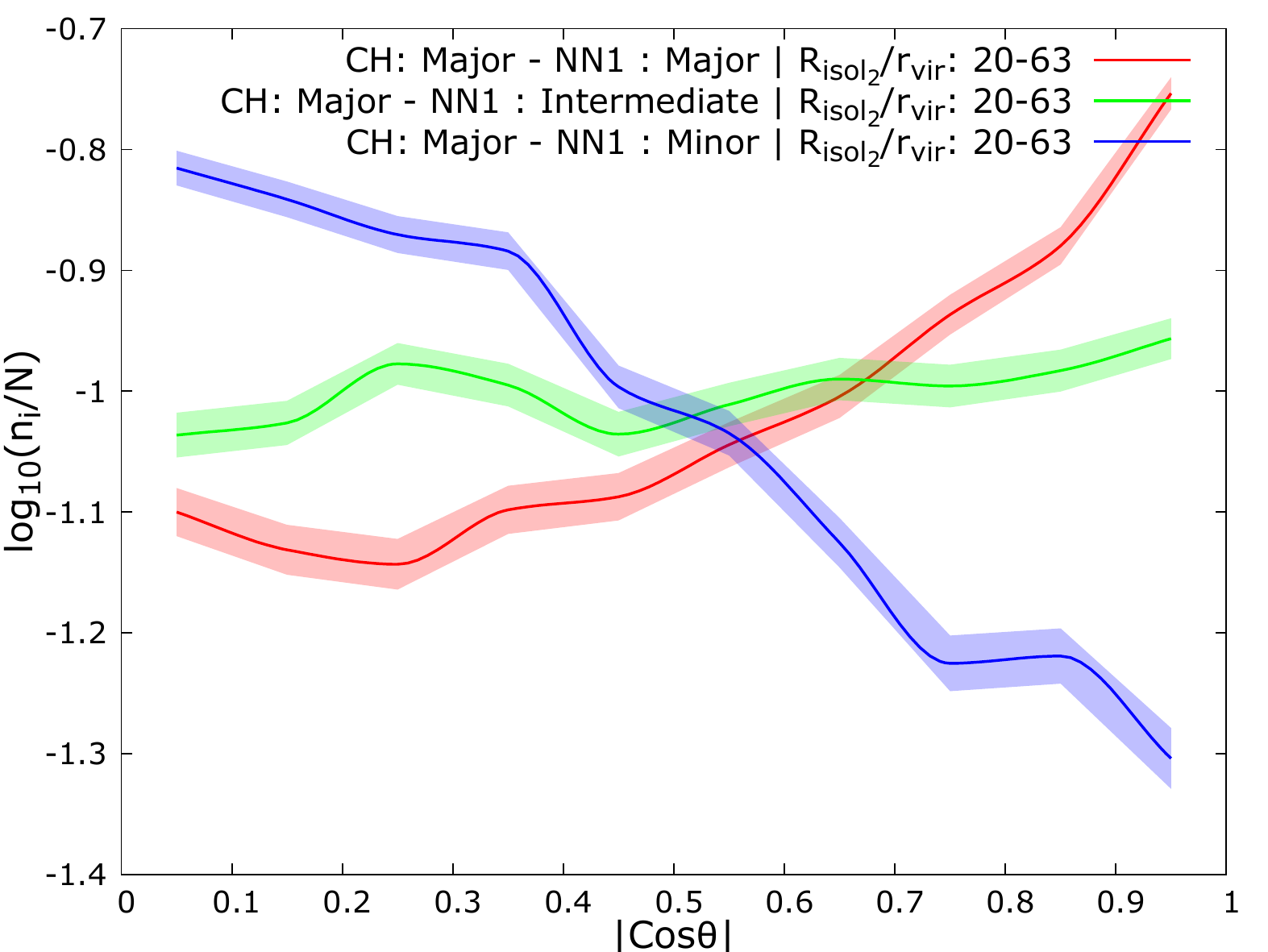}
\caption{Frequency distribution of $|\cos(\theta)|$, where $\theta$ is the angle between the major axis of the CH and the major (red), intermediate (green), and minor (blue) axis of its first nearest neighbor (NN1). Only close halo pair (i.e., CHs separated from their NN1 by $1<R_{\rm isol}/r_{\rm vir}<4$ are considered for this figure. {\em Upper panel:} Close halo pairs where the CHs are separated from their NN2 by $1<R_{\rm isol_{2}}/r_{\rm vir}<4$. {\em Lower panel:} Close halo pairs where the CHs are separated from their NN2 by $20<R_{\rm isol_{2}}/r_{\rm vir}<63$.}
\label{maj-maj}
\end{figure}

 In Fig. \ref{maj-maj}, we present the normalized frequency distributions of $|\cos(\theta)|$ for the angles formed by the major axis of the CH and the major (red), intermediate (green), and minor (blue) axis of its NN1. The results are presented for the extreme cases of pair isolation: pairs in dense environments with $1<R_{\rm isol_{2}}/r_{\rm vir}<4$ (upper panel) and extremely isolated pairs with $20<R_{\rm isol_{2}}/r_{\rm vir}<63$ (lower panel) (the specific width of the latter separation bin is selected to avoid small-number statistics). For both isolation cases, although they are significantly stronger for the latter, we find the distributions (in red) increasing with $|\cos{\theta}|$ for the halo pair's major axes, thereby showing a clear tendency to align. This is accompanied (see blue distributions), as expected, by a tendency of misalignment between the major axis of the CH and the minor axis of the NN1. The green distributions appear to show a weak tendency of alignment between the major axis of CHs and the intermediate axis of their NN1. The significantly more pronounced major axes alignment of isolated pairs (red, lower panel) compared to pairs found in dense environments (red, upper panel) indicate that the original anisotropic collapse of matter along the same large-scale filaments, which could be the cause of strong major-axes alignments of neighboring halos, survive in isolated regions while they get disrupted by strong tidal forces in dense environments.

It is worth investigating whether shape-shape alignments depend on whether the CHs are prolate or oblate. To this end, in Section \ref{Pro_vs_Obl}, we present how we repeated our analysis separately for prolate CHs (Population A) with prolate NN1 and for oblate CHs (Population B) with prolate NN1. We do not analyze CHs with oblate NN1 due to the small number of neighbors.

\subsubsection{Prolate versus oblate CHs}\label{Pro_vs_Obl}
\begin{figure}
\centering
\includegraphics[width=0.48\textwidth]{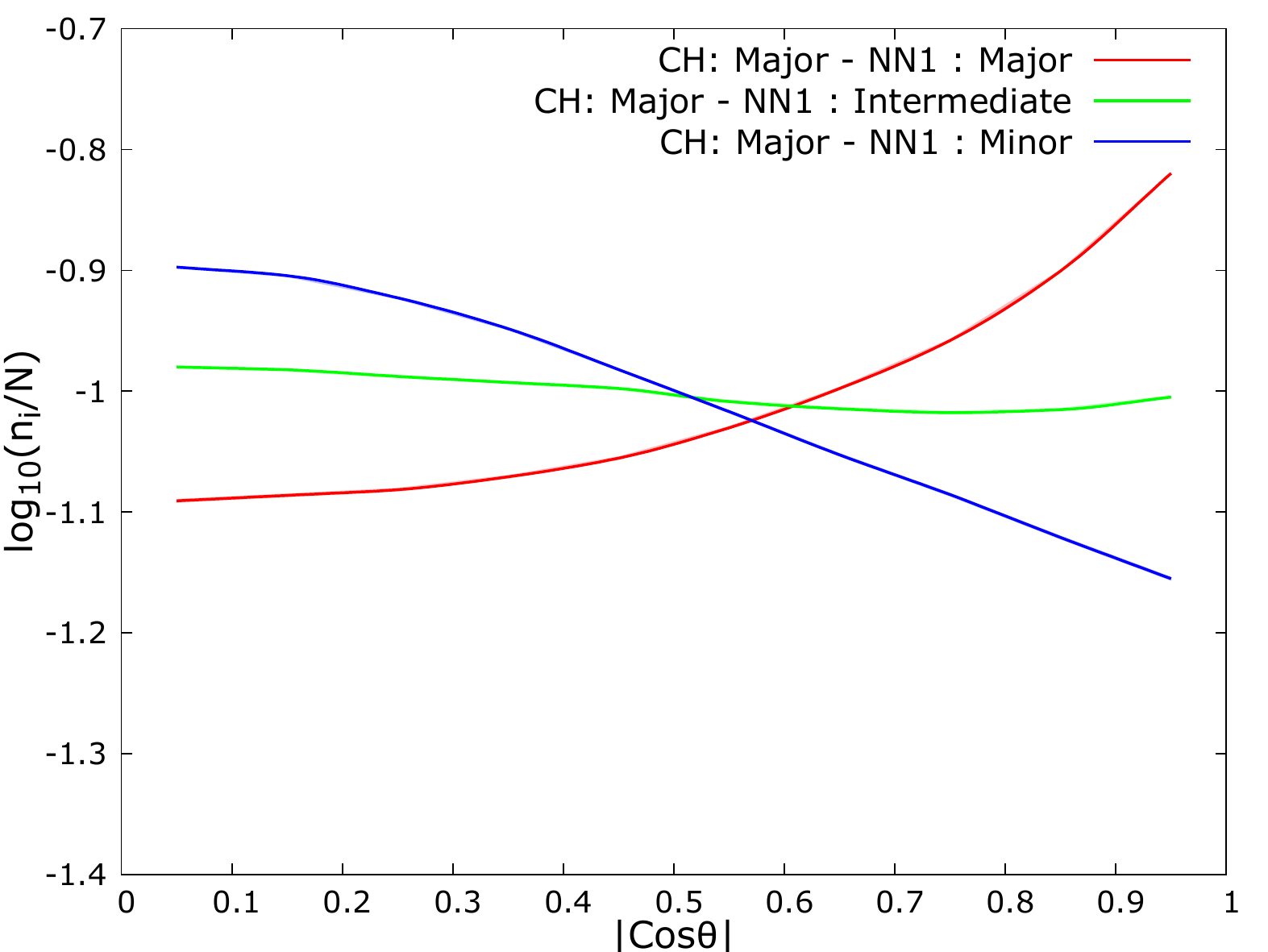}\hfill\\
\includegraphics[width=0.48\textwidth]{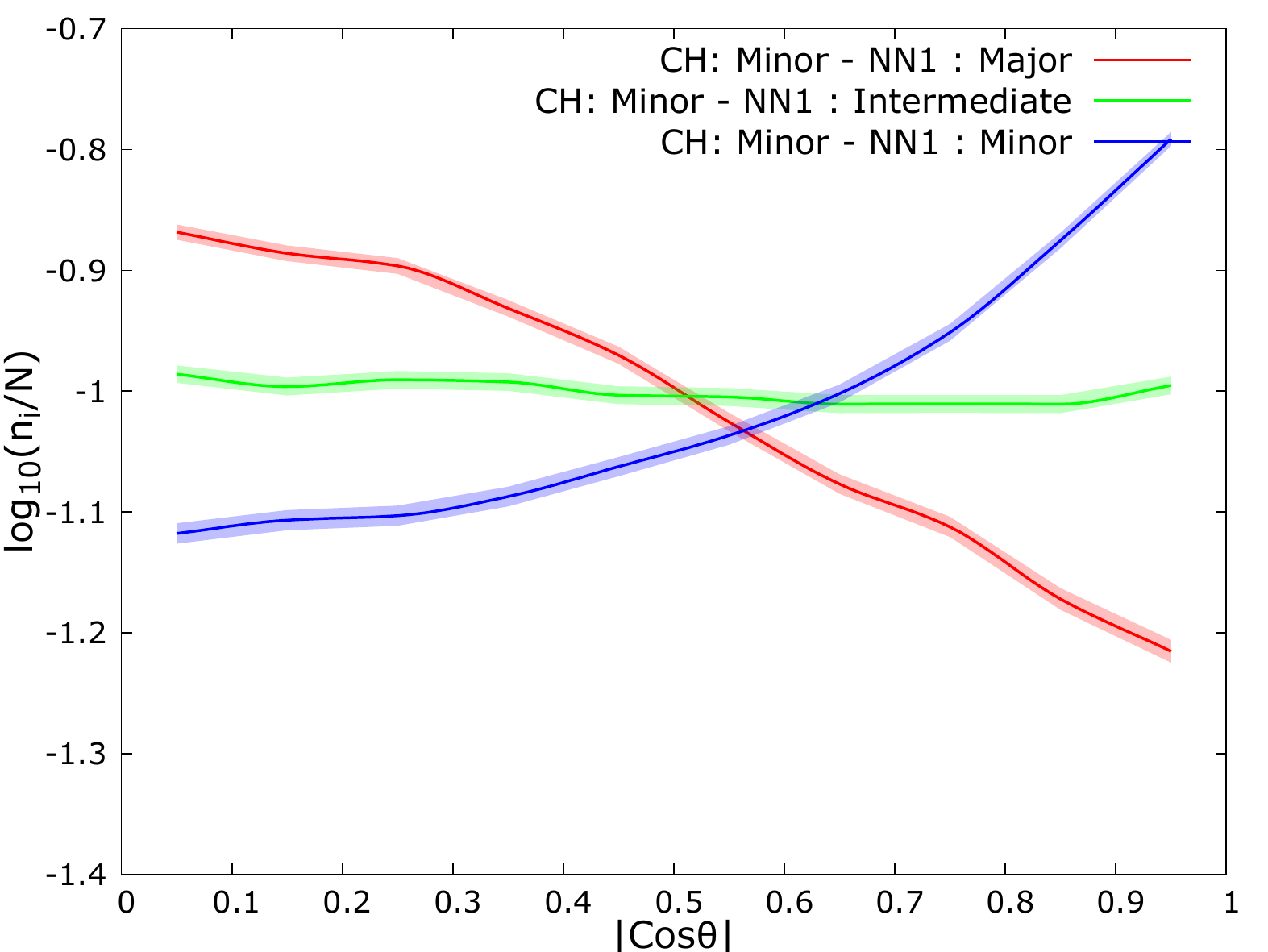}
\caption{Frequency distributions of $|\cos{\theta}|$. Upper panel: $\theta$ is the angle between the major axis of the prolate CHs (Population A) and the major (red), intermediate (green), and minor (blue) axis of their prolate NN1. Lower panel: $\theta$ is the angle between the minor axis of the oblate CHs (Population B) and the major (red), intermediate (green), and minor (blue) axis of their prolate NN1.}
\label{CH-ProvsObl}
\end{figure}

 In Fig. \ref{CH-ProvsObl}, we present the frequency distribution of $|\cos{\theta}|$, where $\theta$ is the angle between the major (minor) axis of prolate (oblate) CHs and the major (red), intermediate (green), minor (blue) axis of their {\bf} prolate NN1. For prolate CHs (upper panel), the results are qualitatively similar to those obtained for the whole sample, that is, the members of our DM halo pairs have preferentially their major axes aligned. For the case of oblate CHs (lower panel),
searching for possible alignments of the minor axis seems more meaningful and, indeed, we find a strong minor-major axis misalignment with their prolate NN1. However, we also find a clear alignment of the minor axis of the oblate CHs with the minor axis of their prolate NN1, having roughly equal minor and intermediate axes. The strong preference for alignment with the minor axis probably indicates the fact that although the prolateness of the majority of halos and the statistical alignment of their shapes is a result of merging and mass accretion in the direction of the filament ridge, the collapse in the perpendicular direction also plays an important role in defining the relative orientation of halos and specifically in aligning the minor axes of the halos, regardless of their shape.

\begin{figure}
\centering
\includegraphics[width=0.48\textwidth]{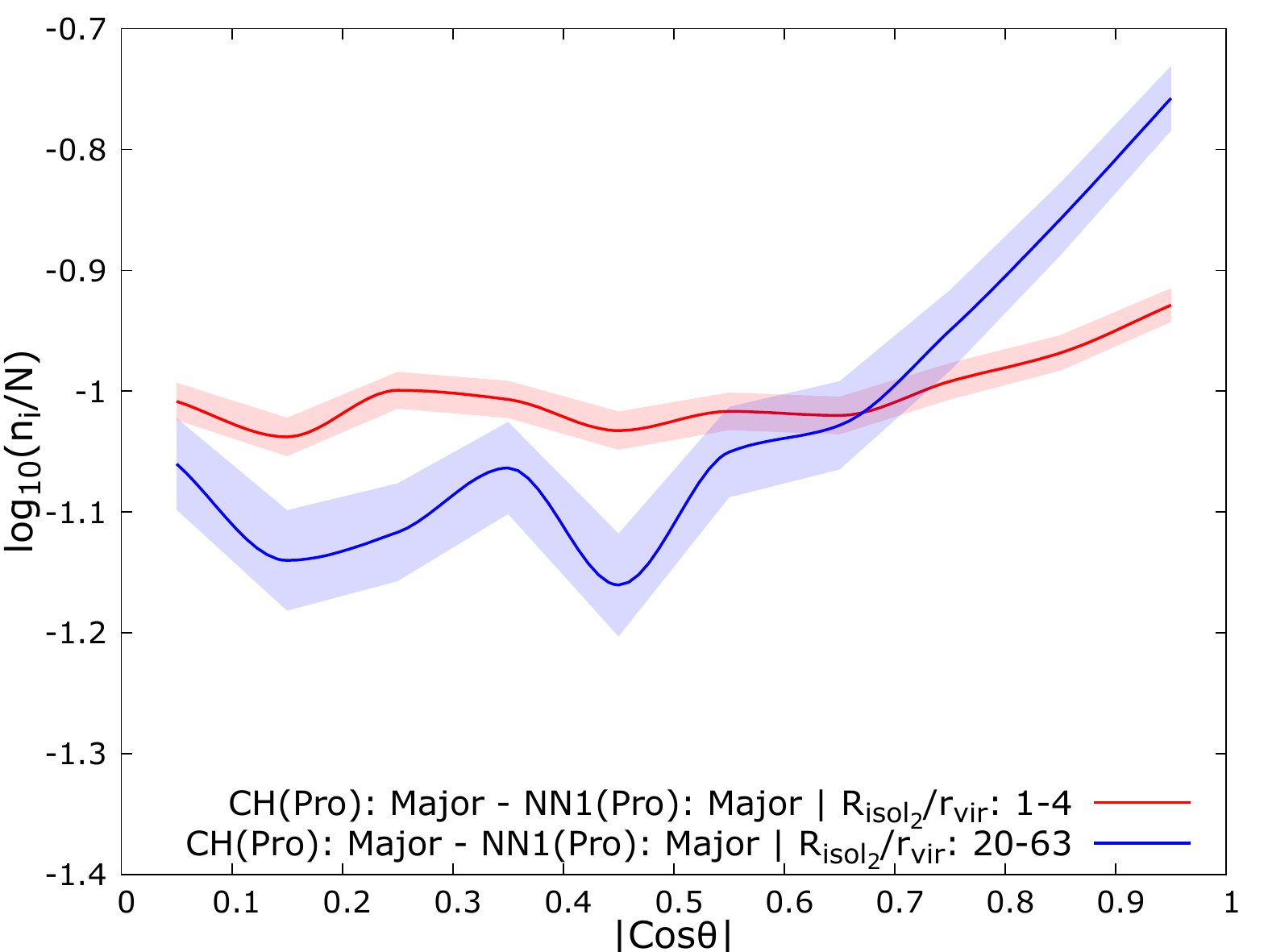}\hfill\\
\includegraphics[width=0.48\textwidth]{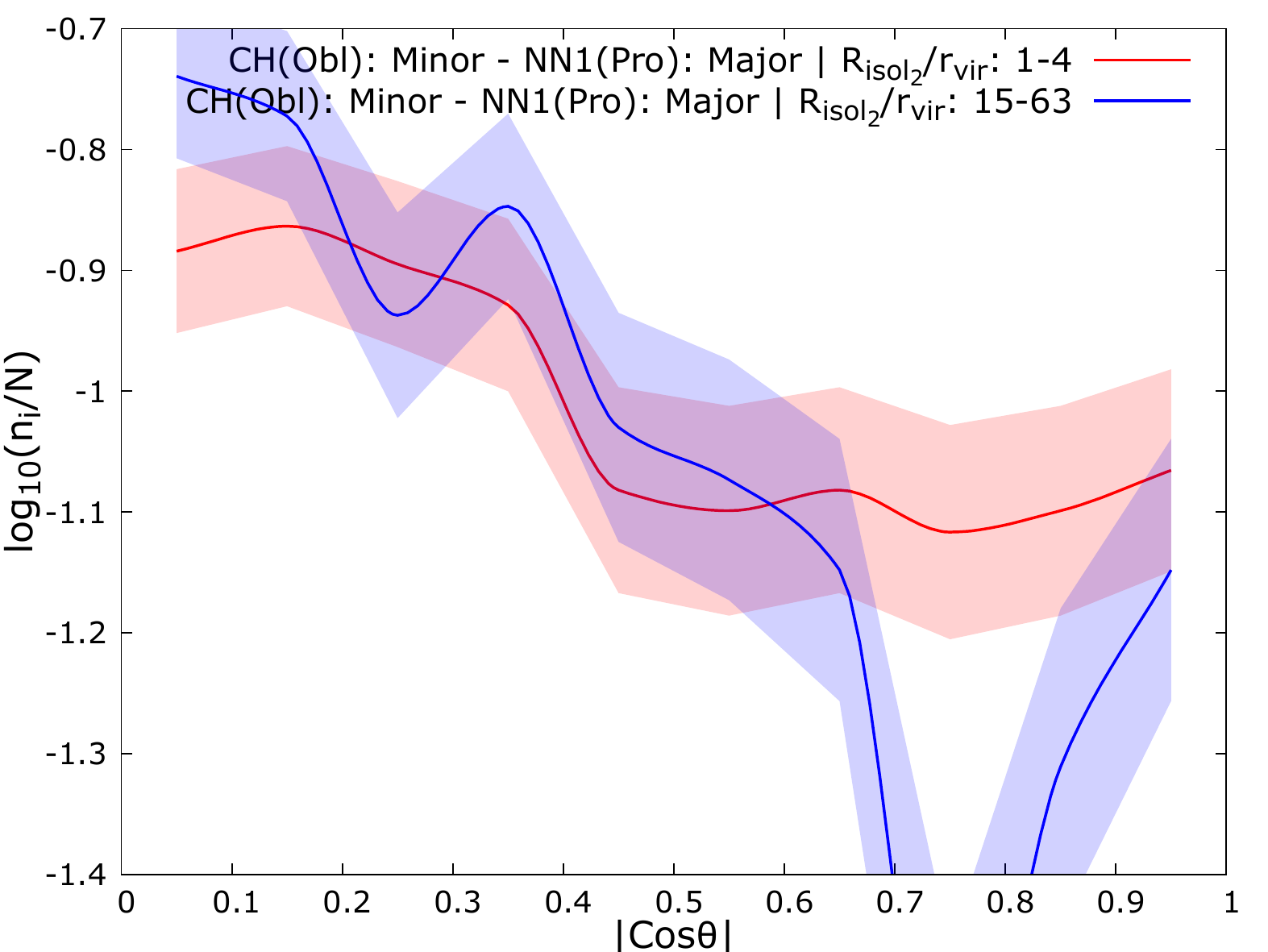}
\caption{Comparison of the shape alignment between the close halo pairs (i.e., CHs separated from their NN1 by $1<R_{\rm isol}/r_{\rm vir}<4$) of two distinct halo populations, based on their shape, for two extreme environments. {\em Upper panel} (Population A): Frequency distribution of $|\cos(\theta)|$, where $\theta$ is the angle between the major axis of the prolate CH and the major axis of its prolate first nearest neighbor (NN1) for two different isolation bins: close halo pairs where the CHs are separated from their NN2 by $1<R_{\rm isol_{2}}/r_{\rm vir}<4$ and close halo pairs where the CHs are separated from their NN2 by $20<R_{\rm isol_{2}}/r_{\rm vir}<63$. {\em Lower panel} (Population B): Frequency distribution of $|\cos(\theta)|$, where $\theta$ is the angle between the minor axis of the Oblate CH and the major axis of its prolate first nearest neighbor (NN1) for two different isolation bins: close halo pairs where the CHs are separated from their NN2 by $1<R_{\rm isol_{2}}/r_{\rm vir}<4$ and close halo pairs where the CHs are separated from their NN2 by $15<R_{\rm isol_{2}}/r_{\rm vir}<63$.}
\label{CH-ProlvsObl-isol}
\end{figure}

In Figure \ref{CH-ProlvsObl-isol}, we explore whether environment has any effect on the shape alignments of Population A and B halos. We find that in low-density environments (see red lines), both tendencies (major-major axes alignments for Population A and minor-major axes alignments for Population B) become more pronounced.

Our results regarding shape alignments of halo pairs, along with the minor axis-spin alignments, support the findings of relevant studies regarding the shape and kinematics alignments of DM halos with respect to the cosmic filaments. The DM halo shapes tend to retain the orientation of the filaments in which they form, regardless of their mass, while the relative direction of their spin with the cosmic web is highly dependent on mass \citeg{GaneshaiahVeena2018}. Thus, our high-mass halos are expected to have spins that are statistically perpendicular to their large-scale structure orientation and, subsequently, perpendicular to their shape orientation as well.

\subsection{Spin-spin alignment}

In this section, we study the alignment of the spin vectors of neighboring DM halos, quantified via $\cos(\theta)\in[-1,1]$, where $\theta \in [0,\pi]$ is the angle between the spin vectors of CHs and their NN1. Since orientation can be defined for spin vectors, in this case, we retain the sign of the $\cos{\theta}$ (see Section \ref{method-alignments}).

\begin{figure}
    \centering
    \includegraphics[width = 0.48\textwidth]{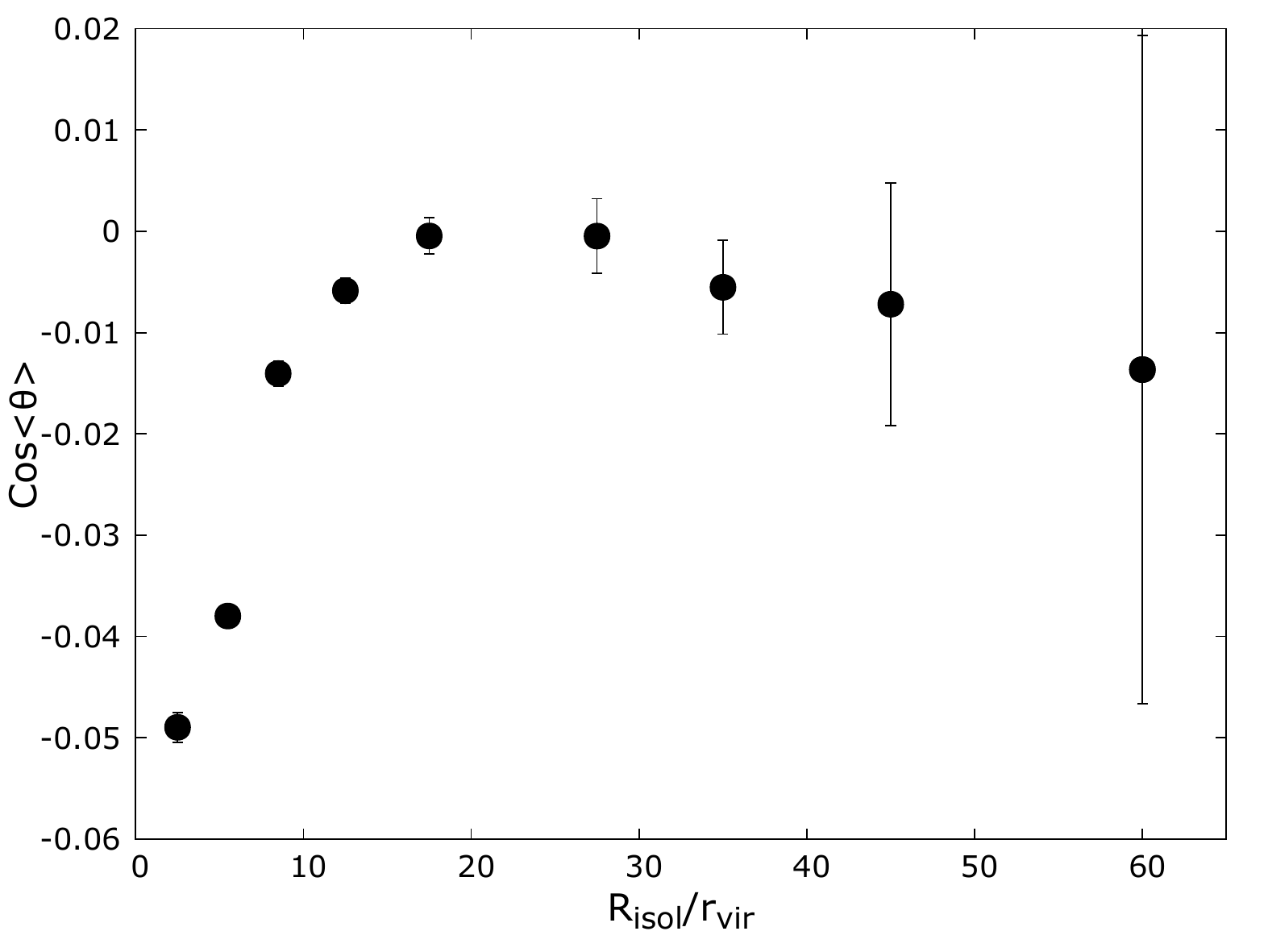}\hfill\\
    \includegraphics[width = 0.5\textwidth]{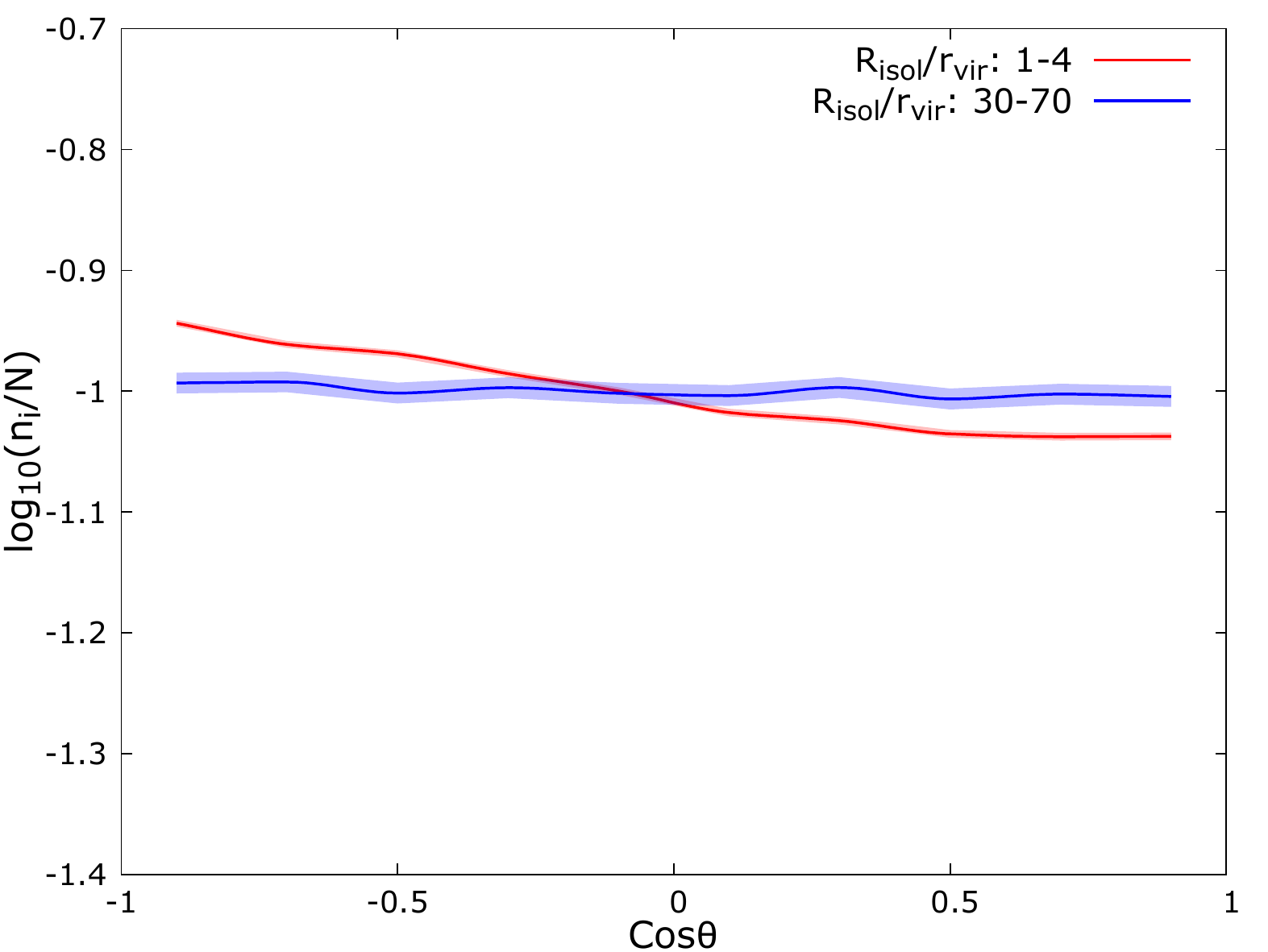}\hfill
    \caption{Relative orientation of the spin vectors of neighboring halos. {\em Upper panel:}  Cosine of the mean angle ($\cos\langle{\theta}\rangle$) between the spin vectors of CHs and their NN1, in bins of $R_{\rm isol}/r_{\rm vir}$. {Lower panel:} Frequency distribution of $\cos{\theta}$ for close neighbors where CHs are separated from their NN1 by $1< R_{\rm isol}/r_{\rm vir}<4$ (red) and for extremely isolated halos where CHs are separated from their NN1 by $30<R_{\rm isol}/r_{\rm vir}<70$ (blue).}
    \label{fig:spins_isol}
\end{figure}

First, we examine the spin alignment of CHs with their NN1 as a function of their separating distance, $R_{\rm isol}$. For this purpose, we present, in the upper panel of Figure \ref{fig:spins_isol},  the cosine of the mean angle ($\cos{\langle\theta\rangle}$) in bins of $R_{\rm isol}/r_{\rm vir}$. We find that $ \cos{\langle\theta\rangle}$ is very close to $0$, meaning that there is no significant preference for a specific relative orientation of the spins of CHs and their nearest neighbor. However, there is a systematic and clear trend of $\cos{\langle\theta\rangle}$ towards slightly negative values (corresponding to more anti-parallel spins) at small isolation radii ($\lesssim 20 R_{\rm isol}/r_{\rm vir}$).

Motivated by this slight indication, in the lower panel of Fig. \ref{fig:spins_isol}, we present the normalized frequency distributions of $\cos{\theta}$ for the two extreme bins of isolation; for the close neighbors (where CHs and their NN1 are separated by $1<R_{\rm isol}/r_{\rm vir}<4$, red) and for extremely isolated neighbors (where CHs and their NN1 are separated by $30<R_{\rm isol}/r_{\rm vir}<70 $, blue). By comparing the two extreme isolation bins, we find that there is a preference for close neighbors to have anti-parallel spin vectors.

This is an effect that shows that with regard to the direction of the angular momentum of neighboring halos, it is the dynamical interactions between close neighbors that dominate over the common halo formation process, or else, halos located in the same local environment would show a tendency for parallel spin vectors.

\begin{figure}
    \centering
    \includegraphics[width = 0.48\textwidth]{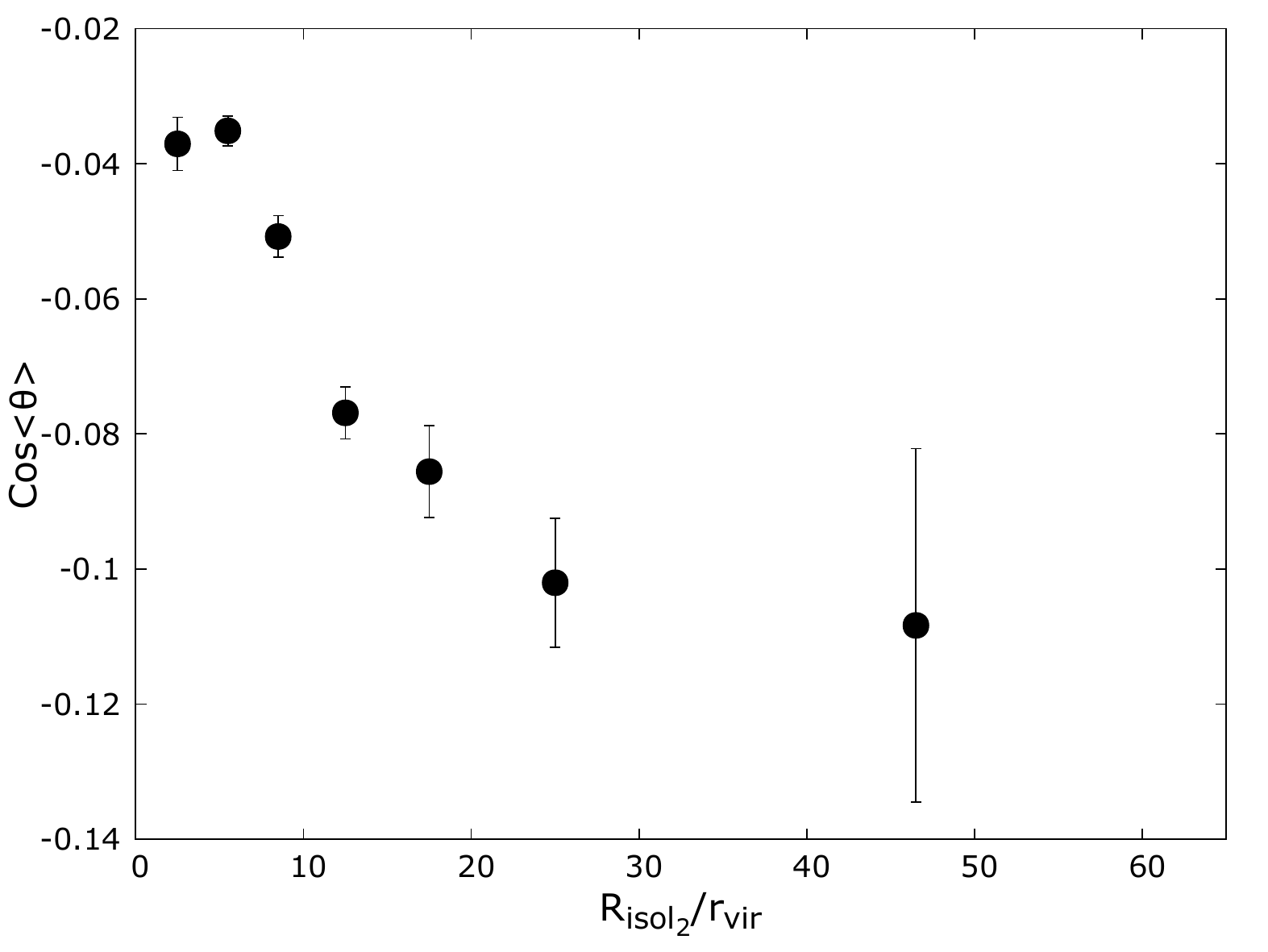}\hfill\\
    \includegraphics[width = 0.48\textwidth]{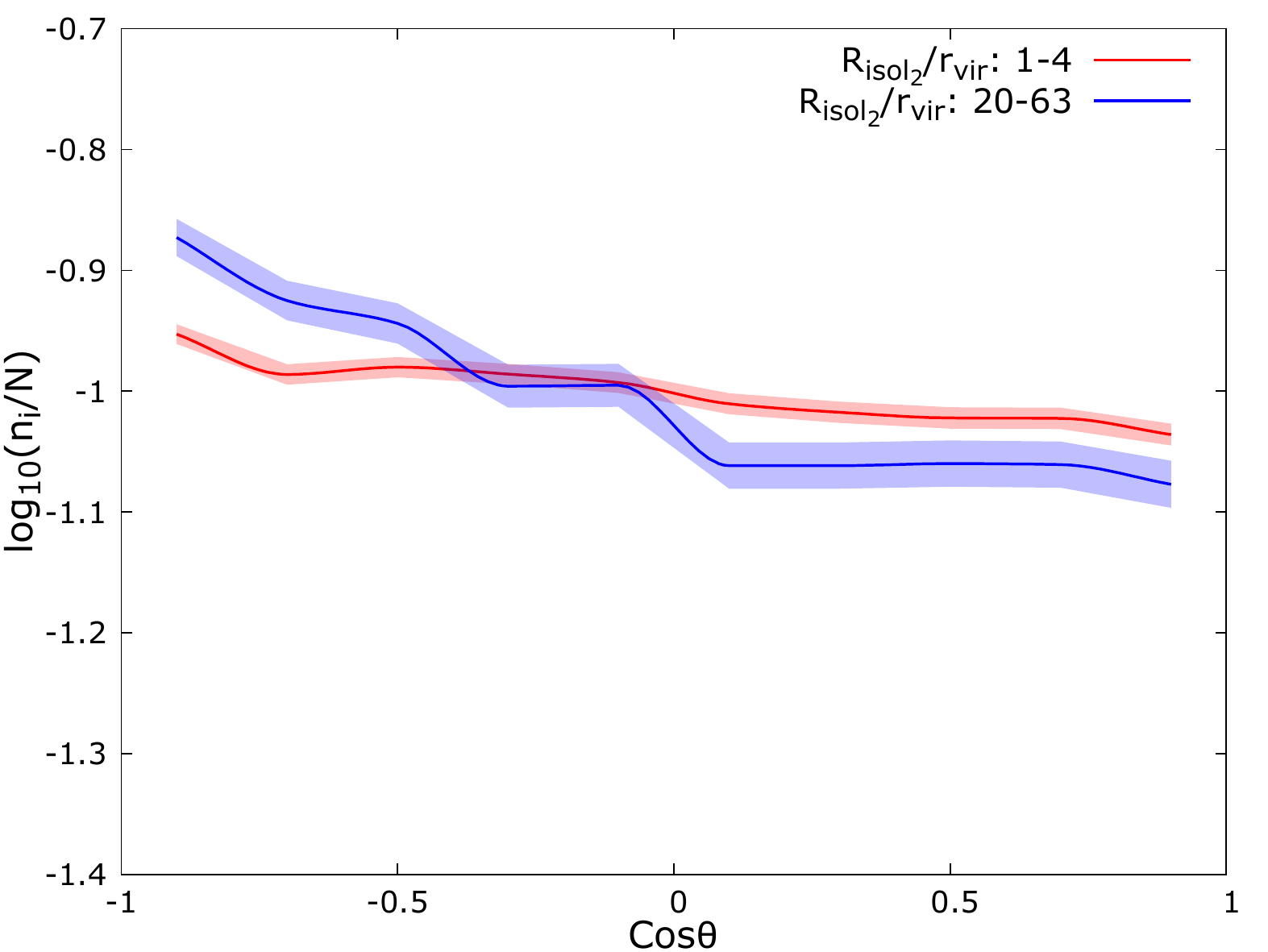}\hfill
    \caption{Relative orientation of the spin vectors of neighboring halos. Only close halo pairs (i.e., CHs separated from their NN1 by $1<R_{\rm isol}/r_{\rm vir}<4$) are considered for this figure. {\em Upper panel:} Cosine of the mean angle ($\cos{<\theta>}$) between the spin vectors of CHs and their NN1 in bins of $R_{\rm isol_{2}}/r_{\rm vir}$. {\em Lower panel:} Frequency distribution of $\cos{\theta}$ for close halo pairs where the CHs are separated from their NN2 by $1<R_{\rm isol_{2}}/r_{\rm vir}<4$ (red) and $20<R_{\rm isol_{2}}/r_{\rm vir}<63$ (blue).}
    \label{fig:spins_isol_pairs}
\end{figure}

We repeat the previous analysis focusing only on close halo neighbors (i.e., halo pairs where CHs and their NN1 are separated by $1<R_{\rm isol}/r_{\rm vir}<4$),
in order to investigate the relative spin orientations in halo pairs as a function of the isolation of the pair, $R_{\rm isol_{2}}/r_{\rm vir}$. In the upper panel of Fig.\ref{fig:spins_isol_pairs}, we present the value of $\langle \cos{\theta}\rangle$ in bins of pair isolation, $R_{\rm isol_{2}}/r_{\rm vir}$, to find that it drops systematically with isolation: the more isolated a halo pair is, the more their spin vectors tend to be anti-parallel. In the lower panel of this figure, we compare the frequency distributions of $\cos{\theta}$ for halo pairs that have a close NN2 ($1<R_{\rm isol_{2}}/r_{\rm vir}<4$, red) and a distant NN2 (20 < $R_{\rm isol_{2}}/r_{\rm vir}< 63$,blue). The stronger anti-parallel alignment tendency of the spin vector for isolated halo pairs supports the notion that dynamical interactions of close neighbors have a dominating effect on the direction of their angular momentum.

A signal of statistically anti-parallel spin vectors of neighboring halos was found by \citet{Hahn2007a}. However, in a follow-up paper, where the authors \citep{Hahn2007b} applied stricter criteria to the exclusion of spurious halos, no specific preference for spin alignments was found. It is important to note that our results, presented in this work, were obtained using halos that comply with the selection criteria of \citet{Hahn2007b} in terms of energy-ratios. Furthermore, we have repeated the above exercise, excluding all CHs and neighbors having $<250$ particles and find an even stronger signal supporting our results. Thus, we are confident that the statistical tendency for anti-parallel spins of close halo pairs is not an artifact induced by spurious halos in our catalog and, rather, it is an interesting and robust result that warrants further investigation.

\begin{figure}
    \centering
    \includegraphics[width = 0.48\textwidth]{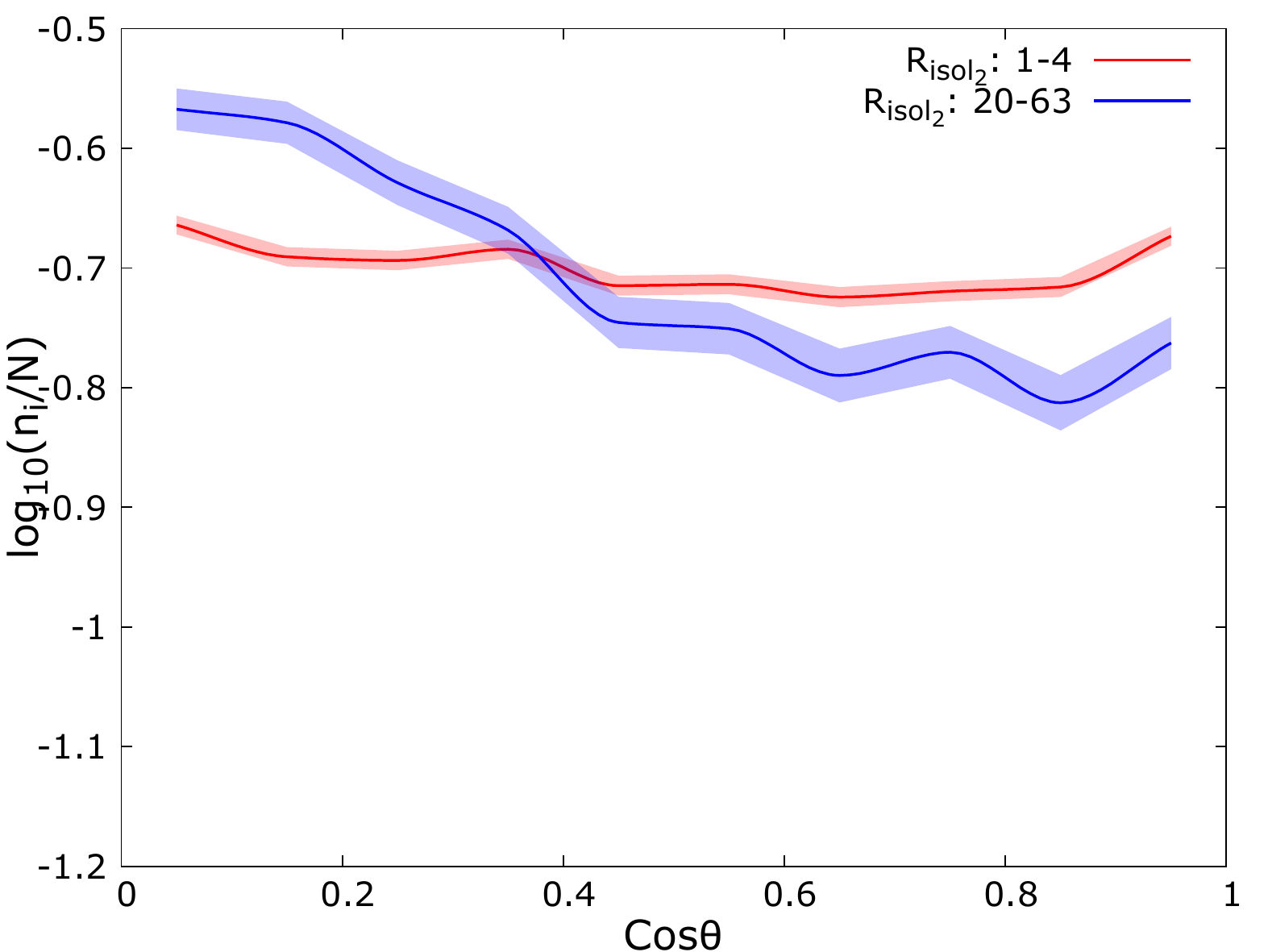}
    \caption{Frequency distributions of $\cos{\theta}$, where $\theta$ is the positional alignment angle between the spin vector and the line defined by the position of the CH and its NN1. Only close halo pairs (i.e., CHs separated from their NN1 by $1<R_{\rm isol}/r_{\rm vir}<4$) in two different in environments are considered for this figure: CHs separated from their NN2 by $1<R_{\rm isol_{2}}/r_{\rm vir}<4$ (red) and  $20<R_{\rm isol_{2}}/r_{\rm vir}<63$ (blue).}
    \label{fig:pos_spin_pairs}
\end{figure}

 In order to have a more complete picture of spin-spin alignments, we study the spin orientation of both CHs and NN1 with respect to the line joining their center of mass. For this purpose, we consider only close halo pairs (i.e., CHs separated from their NN1 by $1<R_{\rm isol}/r_{\rm vir}<4$). In Figure \ref{fig:pos_spin_pairs} we present the frequency distributions of the cosine of the positional alignment angle, $\theta\in[0,\pi/2]$, between the spin vector of each halo (CH and NN1) and the line joining their centers of mass, for halo pairs that have a close NN2 ($1<R_{\rm isol_{2}}/r_{\rm vir}<4$, red) and a distant NN2 (20 < $R_{\rm isol_{2}}/r_{\rm vir}< 63$, blue). Our results show that for the case of extremely isolated halo pairs there is a preference for their spins to be perpendicular with respect to the line connecting the halo pair.

\subsection{Redshift evolution}

The light-cone halo catalog we use throughout this work renders it possible to study the dependence of our results on redshift. Thus, we repeated our analysis in two roughly equal-volume redshift-bins (i.e., $z<0.453$ and $0.453\le z<0.65$) to find that although the mean values of the parameters depend on redshift, as expected from halo evolution, with higher-redshift halos being more aspherical, prolate, less virialized, and displaying higher spins, the overall dependencies of halo-properties on the environment, as well as their correlations, do not change within our redshift range.

\section{Conclusions}
In this work, we investigate how tje properties of DM halos from the DEUS $\Lambda$CDM light-cone simulation depend on halo environment, quantified by the normalized isolation radius, $R_{\rm isol}/r_{\rm vir}$. Moreover, we investigated if these properties are correlated and to what extent these correlations are affected by the local environment. We summarize our results below:

\subsection{Dependence of DM halo dynamical properties (halo shape, spin, and virial state) on isolation}
\begin{enumerate}
      \item DM halo shape depends strongly on isolation. Specifically, we find that isolated halos tend to be more aspehrical and prolate, which is in agreement with the results of previous studies. It appears that the shape of DM halos, being prolate-like in their majority, reflects the anisotropic large-scale environment in which they formed. The prolateness of DM halos is more pronounced in quiet (isolated) environments, which is likely due to a lack of the sort of violent interactions and mergers that are typical of high-density regions.
      \item DM halo spin decreases with isolation, likely due to the fact that strong tidal fields are encountered mostly in dense environments.
      \item We find a systematic dependence of halo virial state on isolation. Specifically, the deviation of $|U/K|$ from the expected value of $2$ for virialized halos decreases with halo isolation. We find that this shows that halos in the most dense environments, while they virialize on shorter timescales than isolated halos, they also often suffer secondary infall and close interactions that may disturb their virial state.
   \end{enumerate}

\subsection{Correlations among DM halo dynamical properties}
Our analysis reveals the existence of strong correlations between the halo properties studied.

\begin{enumerate}
    \item More aspherical halos have higher spin values, regardless of the environment, although spin amplitude is systematically higher in dense environments.
    This indicates that it is likely the same physical process determining halo spin acquisition that is at work in both low and high- density environments -- albiet more efficiently in the latter.

    \item {Virialization and spin appear uncorrelated for a wide range of spin values $(0.01<\lambda<0.1)$, corresponding to the large majority of halos, although for $\lambda>0.1$, there seems to be a correlation.
    For halos in denser environments, high spins correspond to less virialized halos (lower values of $|U/K|$), as expected from the fact that in such environments strong interactions and merging takes place constantly. For extremely isolated halos, higher spin corresponds to more virialized halos, which suggests that quieter environments give halos a better chance at virialize. However, inference based on extreme-spin halos should be approached with caution since this population could be possibly contaminated by spurious halos.}

   \end{enumerate}

\subsection{ Spin-shape alignment of DM halos}
Regardless of the halo environment, we find a strong tendency for halo spin vector to orient perpendicular to the halo major axis and to be equally likely to be oriented along the halo intermediate and minor axes. This indicates that the cause of these alignments originates from the local halo formation processes and is independent of the large-scale structure. Since the dominant halo shape is prolate-like, the alignments for prolate CHs are identical to those of the overall halo sample, while for oblate CHs, the spin vector is predominantly oriented along the minor axis and equally likely to be oriented along the major and intermediate axes - which is consistent with the fact that for oblate halos, the major and intermediate axes are statistically equivalent.

\subsection{Shape-shape alignment of neighboring DM halos}
We find that the major axes of neighboring halos are statistically aligned, with a stronger alignment observed in denser environments. Moreover, when we analyze CHs based on the value of $T$, we  find that oblate-like CHs align with the minor axis of their prolate-like NN. This result reveals that although the accretion of matter along filaments plays a significant role in forming prolate-like halo shapes aligned with the filament, the collapse of matter perpendicularly to the filament-ridge determines the minor axis orientation of halos, regardless of their shape. Finally, for the case of close halo pairs, shape alignments become more noticeable when a halo pair is more isolated, indicating that in quieter tidal field regions, halo neighbors retain the shape orientation as determined by their common large-scale formation environment.

\subsection{Spin-spin alignment of neighboring DM halos}
We find a slight but statistically significant preference for anti-parallel spin vectors of close DM halo neighbors. This preference increases with pair-isolation and becomes more evident for the case of extremely isolated halo pairs, indicating that it originates from the initial angular momentum acquisition mechanisms of halos, since interaction and major mergers are minimal in such low-density environments. We also find that halo spins have a a greater tendency to be perpendicular to the line connecting close halo pairs in isolated environments.
We have verified that this result is not due to the mass resolution of our halos.


\begin{acknowledgements}
 This research is partially co-financed by Greece and the European Union (European Social Fund- ESF) through the Operational Programme "Human Resources Development, Education and Lifelong Learning" in the context of the project "Strengthening Human Resources Research Potential via Doctorate Research" (MIS-5000432), implemented by the State Scholarships Foundation (IKY). Authors thank Pier-Stefano Corasaniti and Yann Rasera at Laboratoire Univers et Th\'{e}ories, Observatoire de Paris, France, for providing the dark matter halo catalogs, and Noam Libeskind, Miguel Aragon-Calvo and Mark Neyrinck for discussions and interesting suggestions.
\end{acknowledgements}


\bibliographystyle{aa}
\bibliography{DYNAMICS}

\end{document}